\documentclass[twocolumn,showpacs,preprintnumbers,amsmath,amssymb]{revtex4}

\usepackage{graphicx}
\usepackage{dcolumn}
\usepackage{bm}
\usepackage{color}

\def\gap{\;\rlap{\lower 2.5pt
 \hbox{$\sim$}}\raise 1.5pt\hbox{$>$}\;}
\def\lap{\;\rlap{\lower 2.5pt
   \hbox{$\sim$}}\raise 1.5pt\hbox{$<$}\;}
\def\gsim{\;\rlap{\lower 2.5pt
 \hbox{$\sim$}}\raise 1.5pt\hbox{$>$}\;}
\def\lsim{\;\rlap{\lower 2.5pt
 \hbox{$\sim$}}\raise 1.5pt\hbox{$<$}\;}
\def\msun{{\rm\,M_\odot}}

\def\cm{{\rm\,cm}}
\def\sec{{\rm\,s}}

\def\sr{{\rm\,sr}}

\def\cm{{\rm\,cm}}

\def\kpc{{\rm\,kpc}}

\def\GeV{{\rm\,GeV}}

\def\sec{{\rm\,s}}
\def\sr{{\rm\,sr}}

\def\spose#1{\hbox to 0pt{#1\hss}}
\def\lta{\mathrel{\spose{\lower 3pt\hbox{$\mathchar''218$}}
     \raise 2.0pt\hbox{$\mathchar''13C$}}}
\def\gta{\mathrel{\spose{\lower 3pt\hbox{$\mathchar''218$}}
     \raise 2.0pt\hbox{$\mathchar''13E$}}}
\newcommand{\beq}{\begin{equation}}
\newcommand{\eeq}{\end{equation}}
\newcommand{\be}{\begin{equation}}
\newcommand{\ee}{\end{equation}}

\newcommand{\ls}{\mathrel{\raise1.16pt\hbox{$<$}\kern-7.0pt 
\lower3.06pt\hbox{{$\scriptstyle \sim$}}}}         
\newcommand{\gs}{\mathrel{\raise1.16pt\hbox{$>$}\kern-7.0pt 
\lower3.06pt\hbox{{$\scriptstyle \sim$}}}}         

\long\def\comment#1{}

\def\msun{M_{\odot}}
\def\fun#1#2{\lower3.6pt\vbox{\baselineskip0pt\lineskip.9pt
  \ialign{$\mathsurround=0pt#1\hfil##\hfil$\crcr#2\crcr\sim\crcr}}}
\def\lap{\mathrel{\mathpalette\fun <}}
\def\gap{\mathrel{\mathpalette\fun >}}
\newcommand{\ba}{\begin{eqnarray}}
\newcommand{\ea}{\end{eqnarray}}

\begin{document}


\title{Multi-messenger constraints on the annihilating dark matter \\
interpretation of the positron excess}

\author{Miguel Pato}
 \email{pato@iap.fr}
\affiliation{
Institut d'Astrophysique de Paris, 98bis bd Arago, 75014, Paris, France
}%
\affiliation{%
Dipartimento di Fisica, Universita' degli Studi di Padova, Via Marzolo 8, I-35131, Padova, Italy
}%
\affiliation{
Universit\'e Paris Diderot-Paris 7, 10 rue Alice Domon et L\'eonie Duquet 75205, Paris, France 
}%

\author{Lidia Pieri}
 \email{pieri@iap.fr}
\affiliation{
Institut d'Astrophysique de Paris, 98bis bd Arago, 75014, Paris, France 
}%

\author{Gianfranco Bertone}
 \email{bertone@iap.fr}
\affiliation{
Institut d'Astrophysique de Paris, UMR7095-CNRS
}%
\affiliation{
Universit\'e Pierre et Marie Curie, 98bis bd Arago, 75014, Paris, France 
}%

\date{\today}

\begin{abstract}
The rise in the energy spectrum of the positron ratio, observed by the PAMELA satellite 
above 10 GeV, and other cosmic ray measurements, have been interpreted as a possible
signature of Dark Matter annihilation in the Galaxy. However, the large number of free parameters, and 
the large astrophysical uncertainties, make it difficult to draw conclusive statements 
about the viability of this scenario. Here, we perform a multi-wavelength,
multi-messenger analysis, that combines in a consistent way the constraints arising 
from different astrophysical observations. We show that if standard assumptions 
are made for the distribution of Dark Matter (we build models on the recent 
Via Lactea II and Aquarius simulations) and the propagation of cosmic rays, 
current Dark Matter models cannot explain the observed positron flux without exceeding the
observed fluxes of antiprotons or gamma-ray and radio photons. To visualize
the multi-messenger constraints, we introduce ``star plots'', a graphical
method that shows in the same plot theoretical predictions
and observational constraints for different messengers and wavelengths.

\end{abstract}

\pacs{95.35.+d, 98.35.Gi, 98.35.Jk}
\maketitle

\section{Introduction}

\par Dark Matter (DM) annihilation or decay can in principle produce significant fluxes of positrons, antiprotons, photons, neutrinos and other secondary particles. Recently, the positron channel has received a lot of attention, since the PAMELA collaboration has released the data relative to the positron fraction \cite{pamela} that exhibit a spectacular rise, which is in agreement with earlier results from AMS-01 \cite{AMS} and HEAT \cite{HEAT,HEAT2,HEAT3}, and compatible with the claimed 300$-$800 GeV excess in the electron plus positron spectrum measured by ATIC-2 balloon flights \cite{atic2}. Lying in the energy range $\gtrsim 10$ GeV, such an excess has prompted a large number of papers putting forward explanations that include DM annihilations or decays in the galaxy, and nearby astrophysical objects like pulsars. 

Here, we consider the DM annihilation hypothesis and perform a multi-messenger analysis in order to constrain the properties of viable DM candidates. 
More specifically, we study positrons, antiprotons, $\gamma$-rays and synchrotron emission due to the propagation of electrons and positrons in the Galactic magnetic field. The multi-messenger approach has already provided useful constraints on DM scenarios. For instance, the non-observation of an excess of cosmic-ray antiprotons up to $\sim$ 100 GeV by the PAMELA satellite \cite{pamelapbar} indicates that hadronic annihilation channels should be strongly suppressed. This motivates a simple distinction between DM candidates: leptophilic, i.e. that annihilate mainly into lepton pairs, and hadrophilic, whose annihilation final states are gauge bosons or quark pairs and induce non-negligible fluxes of both positrons and antiprotons. Obviously, the former override $\bar{p}$ bounds, while the latter need to be rather heavy in order to suppress $\bar{p}$ fluxes below $\sim 100$ GeV, but predict inevitably larger fluxes at higher energies, that should be soon probed. Other interesting messengers for DM searches are high-energy  neutrinos. In fact, neutrino observatories such as Super-Kamiokande \cite{SuperK} and IceCube \cite{IceCube} are able to detect upward-going muons produced in the interaction of high-energy neutrinos within the Earth interior. Therefore, once the DM profile is fixed, neutrino observations of the Galactic Centre region effectively constrain the properties of dark matter, especially for multi-TeV candidates $-$ see e.g.~\cite{Hisanonu,Spolyar:2009kx}.

\par Alternative detection strategies include searches for new physics in accelerators, and direct detection experiments, searching for keV nuclei recoils due to dark matter scattering off nuclei (see e.g.~Refs.~\cite{reviews}). A convincing identification of DM can probably be achieved only through a combination of different detection strategies.

\vspace{0.5cm}

\par Now, if one assumes that DM candidates are thermal relics from the early universe, the present annihilation cross section needs to be $\sigma_{ann}v \sim \mathcal{O}(10^{-26})\textrm{ cm}^3\textrm{s}^{-1}$ in order to produce the observed relic abundance   $\Omega_{cdm}h^2\sim 0.1$. Such a thermal relic with a TeV mass needs a $\sim 10^{3}$ boost in the annihilation flux to accommodate the PAMELA excess \cite{cirelli1}, which can hardly be provided by the ``clumpiness'' of the Galactic halo (see below for a thorough discussion). Non-thermal relics, non-standard cosmologies, or velocity dependent (``Sommerfeld enhanced'') annihilation cross-sections, have been invoked to circumvent this problem, but the question remains of whether the large cross-sections needed to explain the positron data can be made consistent with other astrophysical observations. We perform here an extensive analysis of the multi-messenger constraints in the framework of the latest high-resolution numerical simulations of a Milky Way like halo. We explore two specific classes of models: leptohilic candidates, inspired in Refs.~\cite{AH,N}, and hadrophilic candidates, with a specific emphasis on models of Ref.~\cite{MDM}. For all these models we compute the flux of: {\it i}) positrons, {\it ii})  anti-protons, {\it iii}) gamma-rays from the Galactic Centre, {\it iv}) gamma-rays from the Galactic halo, and {\it v}) synchrotron emission due to the propagation of electrons and positrons at the Galactic Centre. We take into account the dependence of the annihilation cross section on the relative velocity. In particular, we calculate the boost factor due to Sommerfeld enhanced substructures, in the
framework of the Via Lactea II \cite{VL2} and Aquarius \cite{Aq} simulations, and we discuss the consequences for 
the gamma-ray flux from DM annihilations in Galactic and extragalactic halos.

\par The paper is organised as follows. In Sec.~\ref{secnbody} we specify our prescriptions for the distribution 
of DM, including the mass-function and concentration of DM clumps. In Sec.~\ref{candidates}, we introduce the 
DM candidates used in the analysis. Sections \ref{secpos} to \ref{secsync} are then devoted to compute fluxes in the different channels under scrutiny: positrons, antiprotons, $\gamma$-rays and synchrotron. The main conclusions are drawn in the last section.

\section{Astrophysical input}\label{secnbody}

\par Over the last few years, N-body simulations have improved considerably, and recently two groups have published the results of high-resolution simulations, Via Lactea II (VL2) \cite{VL2} and Aquarius (Aq) \cite{Aq}. In the former, both smooth and clumpy components are well fitted by Navarro-Frenk-White (NFW) profiles and the abundance of subhalos follows the rather steep behaviour $M^{-2}$, while in the latter the density profiles seem to be Einasto-like and a shallower subhalo abundance $\propto M^{-1.9}$ is found. A common feature is the presence of many resolved subhalos and a characteristic dependence of their concentration on the position inside the halo, as we will soon explain.

\par The smooth density profile is well modelled, in the VL2 and Aquarius scenarios, by:
\begin{align*}
\quad \rho^{VL2}_{sm}(r) =\frac{\rho_s}{\frac{r}{r_s}\left(1+\frac{r}{r_s}\right)^2} ,
\end{align*}
\vspace{-0.5cm}
\begin{align*}
 \quad \rho^{Aq}_{sm}(r)=\rho_s \, \textrm{exp}\left[-\frac{2}{\alpha}\left(\left(\frac{r}{r_s}\right)^{\alpha}-1\right)\right] , \alpha=0.17 ,
\end{align*}
being $r$ the distance to the galactic centre. The local density is $\rho_{\odot}\equiv \rho_{sm}(r_{\odot})$ with $r_{\odot}=8 \textrm{ kpc}$. Following \cite{lavalle}, the NFW profile of Via Lactea II is considered to saturate at a radius $r_{sat}$ such that $\rho_{sm}(r_{sat})\equiv \rho_{sat}=2\cdot 10^{18} \textrm{ M}_{\odot}\textrm{kpc}^{-3}$.

\par The density profiles inside clumps are NFW in Via Lactea II and Einasto with $\alpha=0.17$ in Aquarius. The corresponding concentration parameters are fitted by a double power law in mass
\begin{eqnarray}
\quad c_{200}(M,r)  &=& \left(\frac{r}{R_{vir}}\right)^{- \alpha_R}  \times  \\ 
&&\left(C1 \left(\frac{M}{\textrm{M}_{\odot}}\right)^{- \alpha_{C1}} + C2 \left(\frac{M}{\textrm{M}_{\odot}}\right)^{- \alpha_{C2}}\right) \nonumber
\end{eqnarray}
where $M$ is the mass of the clump and $r$ again the galactocentric distance. For NFW and Einasto profiles, one has $r_{s}=r_{200}/c_{200}$ and $\rho_s=M_{200}/\left( \int_0^{r_{200}}{dr' \, 4\pi r'^2 f(r')}\right)$ being $r'$ the radial coordinate inside the clump, $f(r')$ the radial dependence of the clump density profile and $r_{200}$ the radius which encloses an average density equal to 200 times the critical density of the universe. In this way, the clump inner density profile $\rho_{cl}(M,r,r')$ is unambiguously defined once the clump mass $M$ and the distance to the galactic centre $r$ are specified. A quantity that turns out to be relevant for galactic positrons and antiprotons is the so-called annihilation volume, $\xi(M,r)=\int_0^{r_{200}}{dr' \, 4\pi r'^2 (\rho_{cl}(M,r,r')}/\rho_{\odot})^2$.

\par Another important input from N-body simulations is the spatial and mass distribution of clumps:
\begin{equation*}
\textrm{VL2}: \quad \frac{d^2N_{sh}}{dMdV}(M,r)=\frac{A_{sh}(M/\textrm{M}_{\odot})^{-2} }{\left(1+\frac{r}{R_a}\right)^2},
\end{equation*}
\vspace{-0.5cm}
\begin{align*}
\textrm{Aq}: & \quad \frac{d^2N_{sh}}{dMdV}(M,r)=A_{sh} (M/\textrm{M}_{\odot})^{-1.9} \times \\
 & \textrm{exp}\left[-\frac{2}{\alpha} \left(\left(\frac{r}{R_{a}}\right)^{\alpha}-1\right)\right] , \alpha=0.678
\end{align*}
in units of $\textrm{ M}_{\odot}^{-1}\textrm{kpc}^{-3}$. In the following, we will refer to these expressions as $\rho_{sh}$.

The normalization $A_{sh}$ is fixed according to the findings of numerical simulations. In the VL2 case, 
we impose that 10 \% of the galaxy mass $M_{vir}$ is virialized in structures with mass in the range $
[10^{-5} M_{vir}, 10^{-2} M_{vir}]$. In the case of Aquarius, we require that 13.2 \% of $M_{vir}$ is concentrated in halos with mass between $1.8 \times 10^{-8} M_{vir}$ (corresponding to the mass resolution in the Aquarius simulation) and $10^{-2} M_{vir}$.

\par
\begin{figure}
  \centering
  \includegraphics[width=8cm,height=8cm]{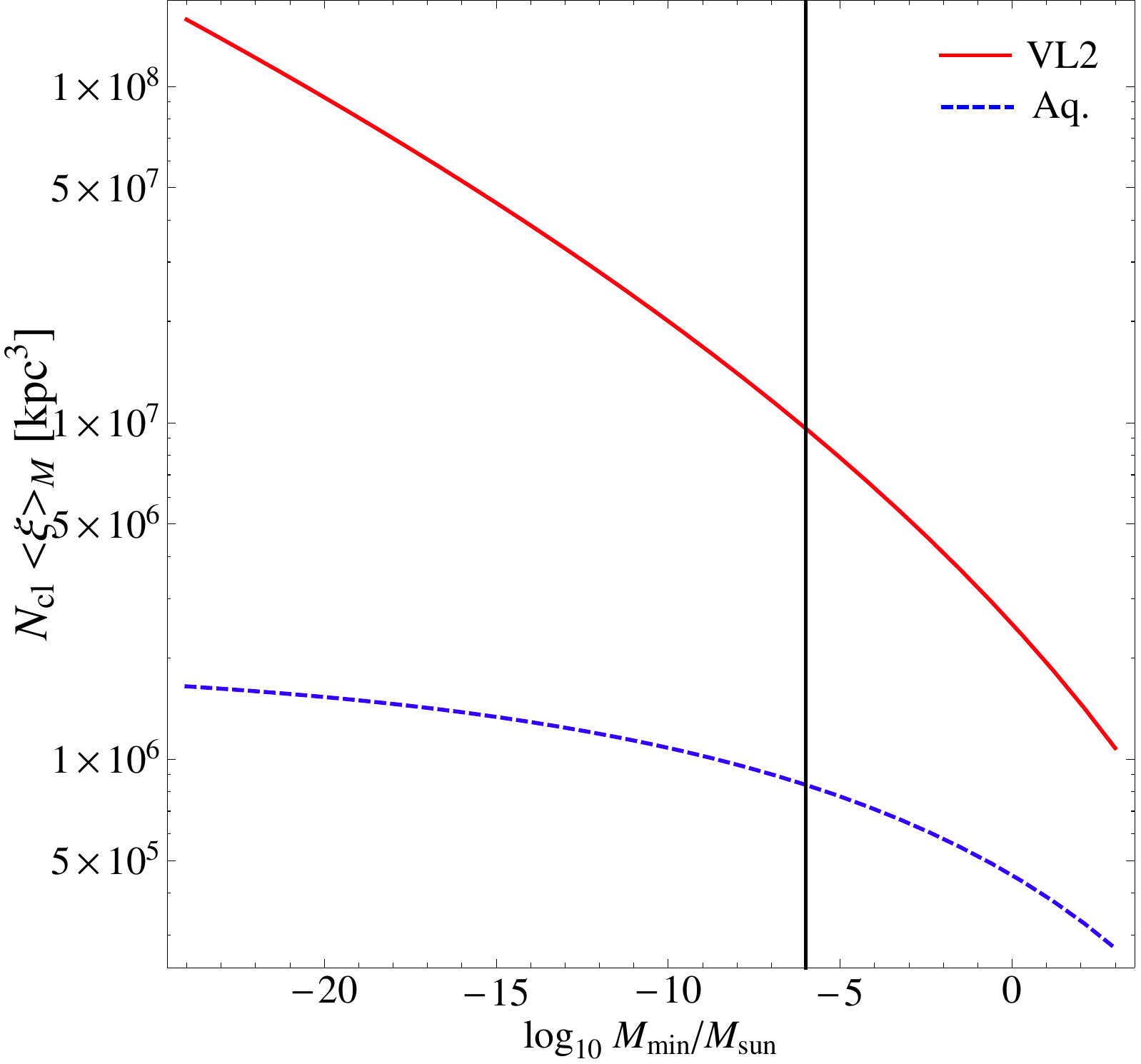}
  \caption{\fontsize{9}{9}\selectfont (Color online) The quantity $N_{cl}\langle \xi \rangle_{M}$ as a function of $M_{min}$ for Via Lactea II (solid red line) and Aquarius (dashed blue line). The vertical line indicates the value used throughout the rest of the present work, $M_{min}=10^{-6} \textrm{ M}_{\odot}$.}\label{figNclxiM}
\end{figure}

 It is convenient to recast the above distribution in the form $\frac{d^2N_{sh}}{dMdV}(M,r)=N_{cl}\frac{dP_M}{dM}(M)\frac{dP_V}{dV}(r)$, where $\int_{M_{min}}^{M_{max}}{dM \, \frac{dP_M}{dM}}=1=\int_0^{R_{vir}}{dr \, 4\pi r^2 \frac{dP_V}{dV} }$. This implies the definition of a subhalo mass range: while $M_{max}$ is usually fixed at $\sim 10^{-2}M_{MW}\sim 10^{10}\textrm{ M}_{\odot}$, $M_{min}$ depends on the nature of dark matter. For supersymmetric neutralinos, for example, it varies in the interval $10^{-12}-10^{-3}\textrm{ M}_{\odot}$ \cite{bringmann}, but it may also be much larger or smaller. We choose to fix $M_{min}=10^{-6}\textrm{ M}_{\odot}$ which is a typical value for WIMPs \cite{hofmann}. As will be shown in sections \ref{secpos} and \ref{secpbar} the positron and antiproton clumpy fluxes scale with the quantity $N_{cl}\langle \xi \rangle_{M}$, where $\langle \xi \rangle_{M}=\int_{M_{min}}^{M_{max}}{dM \, \frac{dP_M}{dM}(M) \xi(M,r_{\odot})}$. Thus, for a given subhalo population (either Via Lactea II or Aquarius) a different value of $M_{min}$ may be accounted for by scaling the $e^{+}$ and $\bar{p}$ clumpy contributions according to figure \ref{figNclxiM}.

\par Other relevant quantities are the total mass in clumps $M_{cl}^{tot}=N_{cl}\int_{M_{min}}^{M_{max}}{dM \, M \frac{dP_M}{dM}}$, the local clump fraction $f_{\odot}=\frac{M_{cl}^{tot}}{\rho_{\odot}}\frac{dP_V}{dV}(r_{\odot})$ and the total clump fraction $f_{cl}^{tot}=\frac{M_{cl}^{tot}}{M_{vir}}$. Table \ref{tab1} displays these and other parameters for both Via Lactea II and Aquarius simulations.

\begin{table}
\centering
\fontsize{9}{9}\selectfont
\begin{tabular}{l|cc}
\hline
\hline
      & Via Lactea II & Aquarius \\
\hline
$R_{vir} \, [\textrm{kpc}]$ & 402 & 433 \\
$M_{vir} \, [\textrm{M}_{\odot}]$ & $1.93 \cdot 10^{12}$ & $2.50 \cdot 10^{12}$ \\
$r_s \, [\textrm{kpc}]$ & 21 & 20 \\
$\rho_s \, [10^6 \msun \kpc^{-3}]$ & 3.7 & 2.4 \\
$\rho_{\odot} \, [\textrm{GeV}\,\textrm{cm}^{-3}]$ & 0.19 & 0.48 \\
\hline
$\alpha_R$ & 0.286 & 0.237 \\
C1 & 119.75 & 232.15 \\
C2 & -85.16 & -181.74 \\
$\alpha_{C1}$ & 0.012 & 0.0146 \\
$\alpha_{C2}$ & 0.026 & 0.008 \\
\hline
$A_{sh} \, [\msun^{-1} \kpc^{-3}]$ & $1.7 \times 10^4$ & 25.86 \\
$R_a [\textrm{kpc}]$ & 21 & 199 \\
$N_{cl}$ & $2.79 \cdot 10^{16}$ & $1.17\cdot 10^{15}$ \\
$\frac{dP_V}{dV}(r_{\odot}) \, [\textrm{kpc}^{-3}]$ & $3.20\cdot 10^{-7}$ & $8.47\cdot10^{-8}$ \\
$\langle \xi \rangle_{M} \, [\textrm{kpc}^3]$ & $3.45\cdot10^{-10}$ & $7.19\cdot10^{-10}$ \\
$M_{cl}^{tot} (< R_{vir}) \, [\textrm{M}_{\odot}]$ & $1.05\cdot10^{12}$ & $4.33\cdot10^{11}$\\
$f_{cl}^{tot} (< R_{vir})$ & 0.54 & 0.18 \\
$f_{\odot}$ & $6.3\cdot10^{-2}$ & $2.7\cdot10^{-3}$ \\
\hline
\end{tabular}

\caption{\fontsize{9}{9}\selectfont Parameters fixing the characteristics of the dark matter distribution as deduced from Via Lactea II and Aquarius results. The three blocks refer to parameters of 1) the smooth galactic halo, 2) the concentration parameter and 3) the galactic subhalo population. Notice that we are setting $M_{min}=10^{-6} \textrm{ M}_{\odot}$ to compute $N_{cl}$, $\langle \xi \rangle_{M}$, $M_{cl}^{tot}$, $f_{\odot}$ and $f_{cl}^{tot}$ .}\label{tab1}

\end{table}

\section{Particle physics input}
\label{candidates}
\par The anomalous positron fraction reported by PAMELA and the electron plus positron excess claimed by ATIC have prompted the interest of the particle physics community,  and motivated the quest for DM models leading to enhanced DM fluxes. Among them, strong emphasis has been put on the Sommerfeld enhancement \cite{Somm}, arising from the presence of an attractive potential, that for low relative velocities leads to a peculiar behaviour of the annihilation cross section $\sigma_{ann} v \propto 1/v$ down to a given $v_{sat}$, below which $\sigma_{ann} v$ saturates. In this scheme, one can have dark matter particles that at chemical decoupling $-$ when $v\sim \mathcal{O}(0.1 c)$ $-$ presented the appropriate annihilation cross section of weak strength, $(\sigma_{ann}v)_0 \sim \mathcal{O}(10^{-26})\textrm{ cm}^3\textrm{s}^{-1}$, and today, in the galactic halo, have a much higher $\sigma_{ann}v $ since the local velocity dispersion is $\beta_{\odot} = v_\odot /c \sim 5\cdot 10^{-4}$. The subscript 0 in $\sigma_{ann} v $ denotes the value of the annihilation cross section without Sommerfeld corrections. The Sommerfeld effect typically leads to small corrections of the annihilation cross section at decoupling \cite{HisanoSommWino}, while boosting significantly local anti-matter fluxes. Notice that the relic abundance of a Sommerfeld-enhanced DM particle may be reduced w.r.t.~the standard case by a factor of order unity: indeed, in Ref.~\cite{HisanoSommWino} the authors have computed the thermal relic abundance of wino-like neutralinos including Sommerfeld corrections and found a reduction of about 50\% compared to the standard calculation.

\begin{table*}
\centering
\fontsize{9}{9}\selectfont
{\small
\fontsize{9}{9}\selectfont
\hfill{}
\begin{tabular}{l|c|c|c|c|c|c|c}
\hline
\hline
label & Ref. & $m_{DM}$/TeV & $m_{\phi,s}$/GeV & $(\sigma_{ann} v)_0/(10^{-26} \textrm{cm}^{3}\textrm{s}^{-1})$ & annihilation channel & $S(v_{\odot})$ & $S_{max}$  \\
\hline
$\square$ AH700 & \cite{AH} & 0.70 & 0.10 ($\phi$) & $3$ & $\phi\phi; \phi\to e^{+}e^{-}$ & 43 & 762  \\
$\bullet \textrm{ }$ NT1 & \cite{N} & 1.00 & 34.0 ($s$) & $3$ & $sa; s\to 97\%aa,3\%b\bar{b}; a\to \mu^{+}\mu^{-} $ & 100 & 100  \\
$\circ \textrm{ }$ NT2 & \cite{N} & 1.20 & 5.60 ($s$) & $3$ & $sa; s\to 95\%aa,5\%\tau \bar{\tau}; a\to \mu^{+}\mu^{-} $ & 100 & 100  \\
$\star \textrm{ }$ $\mu\mu$ & \cite{bergstrom09} & 1.60 & $-$ & 3 & $\mu^+\mu^-$ & 1100 & 1100 \\
$\blacksquare$ $\tau\tau$ & \cite{cirelliTANGO} & 2.00 & $-$ & 3 & $\tau^+\tau^-$ & 1000 & 1000 \\ 
\hline
$\times$ MDM3 & \cite{MDM} & 2.70 & $-$ & $\sim 1$ & $WW,ZZ$ & 273 & 273  \\
$\otimes$ MDM5 & \cite{MDM} & 9.60 & $-$ & $\sim 1$ & $WW,ZZ$ & 1210 & 1210  \\
\hline
\end{tabular} }
\hfill{}
\caption{\fontsize{9}{9}\selectfont Properties of DM candidates recently proposed in the literature and presenting Sommerfeld-enhanced cross sections. The label of each model is preceded by the corresponding symbol to be used in the plots of the next sections.}\label{tab2}

\end{table*}

\par The Sommerfeld enhancement is rather model-dependent and thus we choose a few examples in the literature, shown in table \ref{tab2}. These do not cover all the possibilities but are meant to be representative benchmarks. We consider specific implementations of Arkani-Hamed et al. \cite{AH} and Nomura \& Thaler \cite{N} models as leptophilic-like candidates. In the former, the dark matter particle annihilates into pairs of scalars or vector bosons $\phi$ which then decay into muons or electrons. We use $m_{\phi}=100$ MeV (that means $\phi$ decays entirely into $e^+e^-$) and put $\alpha=\lambda^2/(4\pi)=0.01$. As for Nomura \& Thaler models, dark matter annihilates into a scalar $s$ and an axion $a$. We implement the two benchmarks of \cite{N} where $s$ decays mainly into a pair of axions but has small branching ratios to $b\bar{b}$ and $\tau \bar{\tau}$ $-$ see table \ref{tab2}. The axion $a$ is assumed to decay entirely into muons. Placing Nomura \& Thaler models on figure 6 of \cite{AH}, one sees that the Sommerfeld enhancement is already saturated at $v=v_{\odot}$ and reads $\sim 100-300$. We set $S(v_{\odot})=S_{max}=100$ and note that taking a different value is equivalent to rescale $(\sigma_{ann}v)_0$ since we are lying in the saturation regime. Lastly, inspired by the recently published electron plus positron spectrum from Fermi \cite{fermi} and HESS \cite{hess}, the authors of Ref.~\cite{bergstrom09} propose an 1.6 TeV particle annihilating into $\mu^+ \mu^-$, which fits well Fermi, HESS and PAMELA data given an enhancement of 1100. Even though such enhancement is not necessarily due to Sommerfeld corrections, we consider this candidate setting $S(v_{\odot})=S_{max}=1100$. Similarly, we also analyse a 2 TeV particle annihilating into $\tau^+ \tau^-$ with $S(v_{\odot})=S_{max}=1000$ \cite{cirelliTANGO}.

\par For hadrophilic candidates we adopt the case of minimal dark matter \cite{MDM}, namely the fermion triplet and quintuplet. All examples displayed in table \ref{tab2} are Majorana fermions.

\vspace{0.5cm}

\par Essential ingredients to proceed further are the energy spectra dN/dE of positrons, antiprotons and photons produced in dark matter annihilations. Arkani-Hamed et al. models feature particles that annihilate in an 1-step cascade into $e^+e^-$; the relevant formulae for the positron spectrum are given in Appendix A of \cite{nomura}. The Nomura \& Thaler cases considered here annihilate mainly in an 1.5-step cascade into $\mu^+\mu^-$ $-$ which is basically half an 1-step cascade and half a 2-step cascade $-$ and the referred appendix gives the necessary expressions for $dN_{e^{+}}/dE_{e^{+}}$. The corrections due to the branching ratios into $b\bar{b}$ or $\tau \bar{\tau}$ are introduced in the following way: from DarkSUSY code version 5.0.4 \cite{darksusy} we take the positron spectra for the mentioned channels and $m_{DM}=10$ GeV (corresponding to $m_{s}=20$ GeV). Then, following \cite{nomura}, we convolute them to give the $dN_{e^+}/dE_{e^+}$ in the centre of mass of the DM annihilations. The same procedure is applied for $dN_{\bar{p}}/dT_{\bar{p}}$ in the $b\bar{b}$ case. With the small branching ratios into $\tau\bar{\tau}$ and $b\bar{b}$ presented in table \ref{tab2}, the positron spectra obtained in this way are very similar to the ones obtained in a pure 1.5-step cascade. However, an important difference is a non-zero yield of antiprotons (that will turn out to be small) and possibly significant $\gamma$-ray production. Lastly, energy spectra from minimal dark matter annihilations are given in reference \cite{MDM} for $x\gtrsim 10^{-4}$. We implement the $e^{+}$ and $\bar{p}$ spectrum of \cite{MDM} down to $x=10^{-4}$ and a flat $dN/dE$ is assumed below that. This feature does not affect our conclusions. Indeed, the main results will depend on $dN_{e^{+}}/dE_{e^{+}}$ for energies above 50 GeV which are within the range of validity of all spectra used. Radio fluxes will depend on the number of electrons and positrons above $\sim0.1-10$ GeV; for minimal dark matter $-$ where we have implemented a flat $dN/dE$ below a given $x$ $-$ the number of $e^{\pm}$ will then be underestimated and our radio fluxes for those candidates are lower bounds.

\section{Positrons}\label{secpos}
\par Unlike neutral particles, positrons produced in the Milky Way undergo different processes that change their direction and energy while crossing the galactic medium. The galactic magnetic fields, for instance, are responsible for deflection and, due to their (poorly known) inhomogeneities, the evolution of a positron can be treated as a random walk with a certain diffusion coefficient $K_{e^{+}}$. Other important phenomena are energy losses through inverse Compton scattering off the cosmic microwave background and starlight and synchrotron emission, which proceed at a space-independent rate $b(E_{e^{+}})\simeq E_{e^{+}}^2/(\textrm{GeV}\cdot\tau_E)$ with $\tau_E\simeq 10^{16}$ s \cite{Delahaye08,Delahaye07}. Neglecting galactic convective winds and diffusive reacceleration, the number density per unit energy $n_{e^{+}}(t,\textbf{x},E_{e^{+}})\equiv\frac{d^2N_{e^{+}}}{dVdE_{e^{+}}}$ follows the diffusion equation \cite{Delahaye08,Delahaye07}
\begin{align}\label{diffe+}
\frac{\partial n_{e^{+}}}{\partial t}-& K_{e^{+}}(E_{e^{+}})\nabla^2 n_{e^{+}} \nonumber \\
 & \qquad -\frac{\partial}{\partial E_{e^{+}}}\left(b(E_{e^{+}})n_{e^{+}}\right)=Q_{e^{+}}(\textbf{x},E_{e^{+}}) \, ,
\end{align}
and we are interested in positrons from annihilations of dark matter particles with mass $m_{DM}$ and density $\rho_{DM}$ corresponding to the source term
\begin{equation}\label{source}
Q_{e^{+}}(\textbf{x},E_{e^{+}})=\frac{1}{2}\left(\frac{\rho_{DM}(\textbf{x})}{m_{DM}}\right)^2 \sum_k{\langle \sigma_{ann} v \rangle_0^k \frac{dN_{e^{+}}^{k}}{dE_{e^{+}}}(E_{e^{+}}) } \, ,
\end{equation}
where the 1/2 factor is valid for Majorana self-annihilating fermions. The fluxes induced by Dirac fermions, bosons or other particles may be obtained from our results by a simple rescaling.

\begin{table}
\centering
\fontsize{9}{9}\selectfont
\begin{tabular}{l|cccc}
\hline
\hline
      & $L$ [kpc] & $K_0$ [kpc$^2$/Myr] & $\delta$ & $V_c$ [km/s]  \\
\hline
M2  & 1 & 0.00595 & 0.55 & $-$ \\
MIN & 1 & 0.0016 & 0.85 & 13.5 \\
MED & 4 & 0.0112 & 0.70 & 12.0 \\
MAX & 15 & 0.0765 & 0.46 & 5.0 \\
\hline
\end{tabular}
\caption{\fontsize{9}{9}\selectfont Sets of propagation parameters yielding maximal, mean and minimal anti-matter fluxes \cite{Delahaye07,Donato04}.}\label{tabprop}
\end{table}

\par The standard approach to solve \eqref{diffe+} for $n_{e^{+}}$ is to assume steady state conditions (i.e. $\partial n_{e^{+}}/\partial t=0$) and adopt a cylindrical diffusion halo with radius $R_{gal}=20$ kpc and a half-thickness $L$ inside which the diffusion coefficient is supposed to be space-independent \cite{Delahaye08,Delahaye07}, $K_{e^{+}}(E_{e^{+}})\simeq K_0 \left(E_{e^{+}}/\textrm{GeV}\right)^\delta$. The half-thickness $L$ extends much further than the half-thickness of the galactic disk $h\simeq 0.1$ kpc and $n_{e^{+}}$ vanishes at the cylinder boundaries since the particles escape to the intergalactic medium. The propagation model is defined by the set of parameters $(L,K_0,\delta)$ which turn out to be loosely constrained by cosmic ray data, namely B/C measurements. Following \cite{Delahaye07,Donato04} we use the sets of parameters labelled M2, MIN, MED and MAX in table \ref{tabprop} that are likely to reflect the propagation uncertainty on dark matter induced anti-matter fluxes. M2 (MIN) is the set that minimises the positron (antiproton) flux. The value of the galactic wind speed will be used in the antiproton analysis while being neglected here.

\par Once the steady state solution is found, the flux of positrons is given by $\phi_{e^{+}}(\textbf{x},E_{e^{+}})=\frac{v_{e^{+}}}{4\pi} n_{e^{+}}(\textbf{x},E_{e^{+}})$. We disregard solar modulation, since it is unimportant for multi-GeV positrons. Following \cite{lavalle,Delahaye08}, the positron flux at Earth due to the smooth dark matter component of the Milky Way is
\begin{align}\label{smooth}
\phi_{e^{+},sm}^0(E)= & \frac{v_{e^{+}}}{4\pi} \frac{1}{b(E)} \frac{1}{2} \left(\frac{\rho_{\odot}}{m_{DM}}\right)^2 \times \nonumber \\
&  \int_{E}^{\infty}{dE_S \, f_{inj}^{e^{+}}(E_S) \, I_{sm}^{e^{+}} \left(\lambda_D(E,E_S)\right)} \, ,
\end{align}
where the sun is at $(x_{\odot},y_{\odot},z_{\odot})=(8,0,0)$ kpc, $v_{e^{+}}/c=\left(1-m_e^2/E^2\right)^{1/2}$ and $f_{inj}^{e^{+}}(E_S)=\sum_k{\langle \sigma_{ann} v \rangle_0^k \frac{dN^{k}_{e^{+}}}{dE_{e^{+}}}(E_S) }$. $\lambda_D(E,E_S)$ is the positron diffusion length from a source energy $E_S$ down to a detection energy $E \leq E_S$ and reads
\begin{equation*}
\lambda_D(E,E_S)=\sqrt{\frac{4K_0\tau_E}{1-\delta}\left(\left(\frac{E}{\textrm{GeV}}\right)^{\delta-1}-\left(\frac{E_S}{\textrm{GeV}}\right)^{\delta-1}\right)} \, .
\end{equation*}
\par $I_{sm}^{e^{+}}(\lambda_D)$ is the dimensionless halo function and is given by
\begin{equation} \label{Ism}
I_{sm}^{e^{+}}(\lambda_D)= \int_{DZ}{d^3 \textbf{x} \,  \left(\frac{\rho_{sm}(\textbf{x})}{\rho_{\odot}}\right)^2 \, G_{\odot}^{e^{+}}(\textbf{x},\lambda_D)} \, ,
\end{equation}
where DZ stands for the cylindrical diffusive zone and $G_{\odot}^{e^+}$ is the Green function evaluated at the solar neighbourhood:
\begin{align*}
G_{\odot}^{e^{+}}(\textbf{x},\lambda_D)= &\frac{1}{\pi \lambda_D^2} \, \textrm{exp}\left(-\frac{(x-x_{\odot})^2+(y-y_{\odot})^2}{\lambda_D^2}\right) \times \\
&  \qquad G_{1D}^{e^{+}}(z,\lambda_D) \, ,
\end{align*}
with $G_{1D}^{e^{+}}$ given in \cite{lavalle} (and references therein) for the limiting cases $L>\lambda_D$ and $L\leq \lambda_D$.

\vspace{0.5cm}
\par The contribution from one single clump is very similar to the smooth one replacing $\rho_{sm}$ with $\rho_{cl}$ in equation \eqref{Ism}. However, we are interested in the signal from a population of subhalos distributed throughout the galaxy in a certain range of masses, say $\frac{d^2N_{sh}}{dMdV}=N_{cl} \frac{dP_M}{dM} \frac{dP_V}{dV}$. Considering every clump a point source and given the local character of the Green function, the mean positron flux from the clumpy dark matter component in the galaxy is \cite{lavalle}
\begin{align}\label{clumpy}
\langle \phi_{e^{+},cl}^0 &\rangle (E)=  \frac{v_{e^{+}}}{4\pi} \frac{1}{b(E)} \frac{1}{2} \left(\frac{\rho_{\odot}}{m_{DM}}\right)^2 N_{cl} \, \langle \xi \rangle_M \times \nonumber \\
 & \, \int_{E}^{\infty}{dE_S f_{inj}^{e^{+}}(E_S) \, \langle G_{\odot}^{e^{+}} \rangle_V \left(\lambda_D(E,E_S)\right)} \, ,
\end{align}
where
\begin{equation*} \label{Icl}
\langle G_{\odot}^{e^{+}} \rangle_V (\lambda_D)= \int_{DZ}{d^3 \textbf{x} \, G_{\odot}^{e^{+}}(\textbf{x},\lambda_D)} \, \frac{dP_V}{dV}(\textbf{x})
\end{equation*}
and $\langle \xi \rangle_M$ was introduced in section \ref{secnbody}. Equation \eqref{clumpy} is valid if the density profile of the clump does not depend on its position within the Milky Way. Therefore, we set $c_{200}(M)\equiv c_{200}(M,r_{\odot})$ which is anyway reasonable for our analysis since multi-GeV positrons detected at the Earth travelled at most a few kpc \cite{Delahaye08}.

\par Next we include Sommerfeld corrections. The smooth contribution $\phi_{e^{+},sm}^0$ will be boosted by $S(v_{\odot})$ given the local origin of high-energy positrons. As far as clumps are concerned, we assume that the whole population of subhalos presents velocity dispersions below $v_{sat}$ which means the clumpy contribution $\langle \phi_{e^{+},cl}^0 \rangle$ will be roughly rescaled by $S_{max}$. Such simplification is conservative in the sense that we maximise the contribution of substructures $-$ indeed, clumps with masses close to $M_{max}=10^{10} \textrm{ M}_{\odot}$ may not be in the saturation regime, but that would lead to an enhancement smaller than $S_{max}$. In this framework and following \cite{lavalle}, the total positron flux at Earth for a specific dark matter candidate and a certain propagation model is
\begin{equation}
\phi_{e^{+}}(E)=(1-f_{\odot})^2 \, S(v_{\odot}) \, \phi_{e^{+},sm}^0 (E) \, + \, S_{max} \, \langle \phi_{e^{+},cl}^0 \rangle (E) \, .
\end{equation}

\begin{figure}
 \centering
 \includegraphics[width=7.5cm,height=7.5cm]{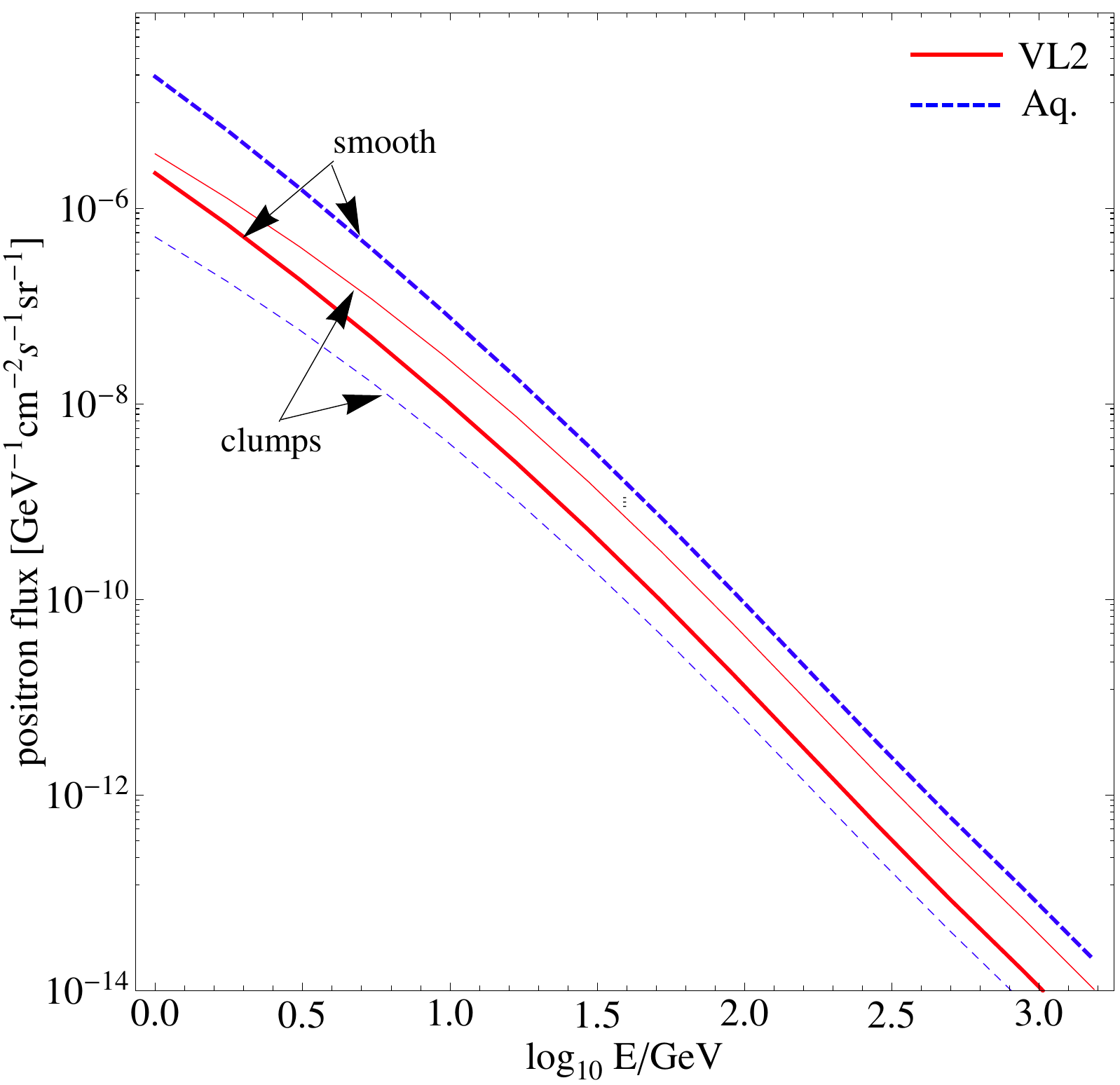}\\
 \includegraphics[width=7.5cm,height=7.5cm]{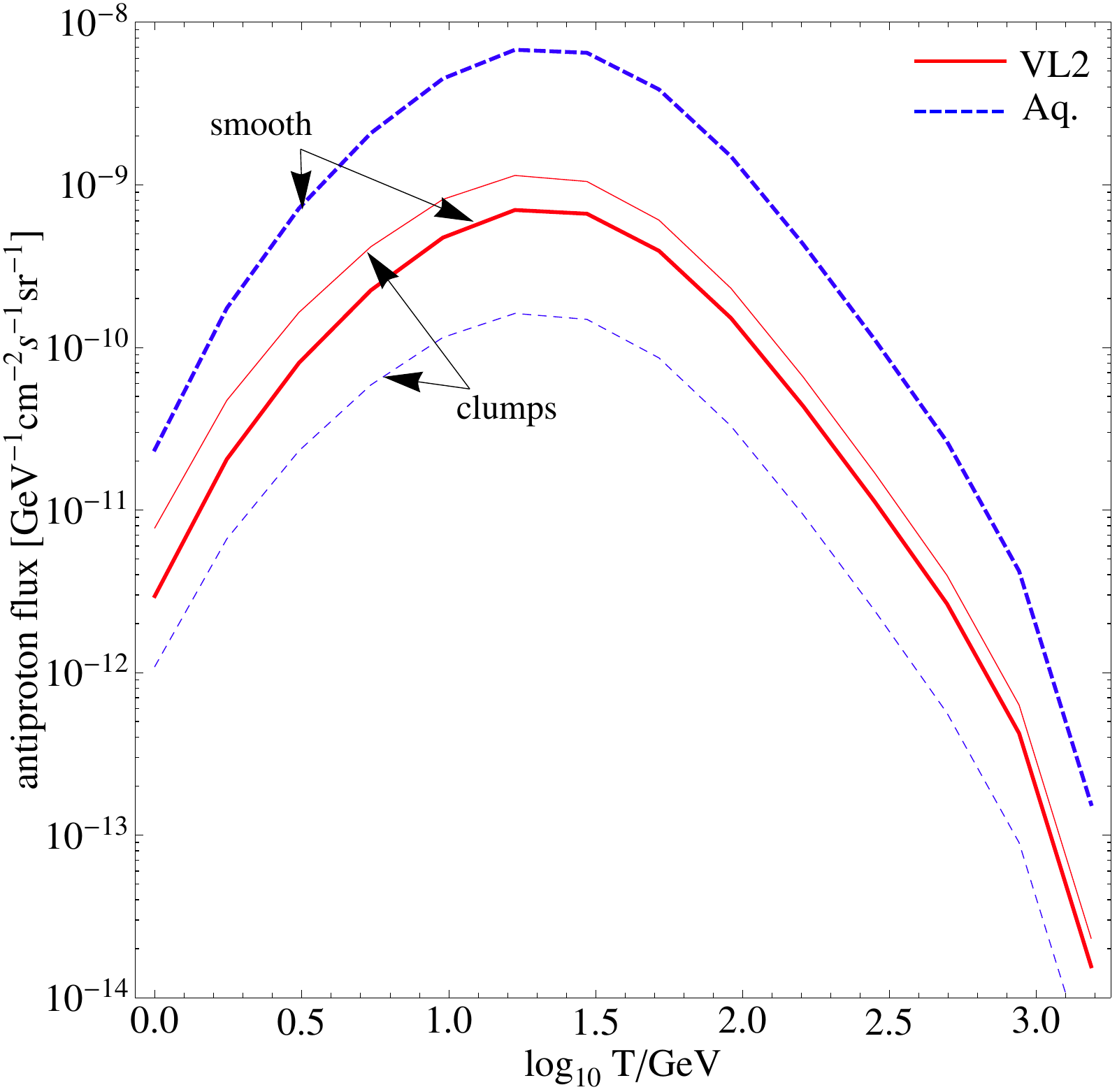}
 \caption{\fontsize{9}{9}\selectfont (Color online) The Sommerfeld-enhanced fluxes $(1-f_{\odot})^2 S(v_{\odot}) \phi_{sm}^0$ and $S_{max}\langle \phi_{cl}^0 \rangle$ of positrons and antiprotons for the MDM3 candidate. Solid red (dashed blue) lines refer to Via Lactea II (Aquarius) density profiles. The thick (thin) curves represent the smooth (clumpy) contribution. The MED propagation model is adopted and $M_{min}=10^{-6}\textrm{ M}_{\odot}$.}
 \label{figLS43}
\end{figure}

\begin{figure}
 \centering
 \includegraphics[width=7.5cm,height=7.5cm]{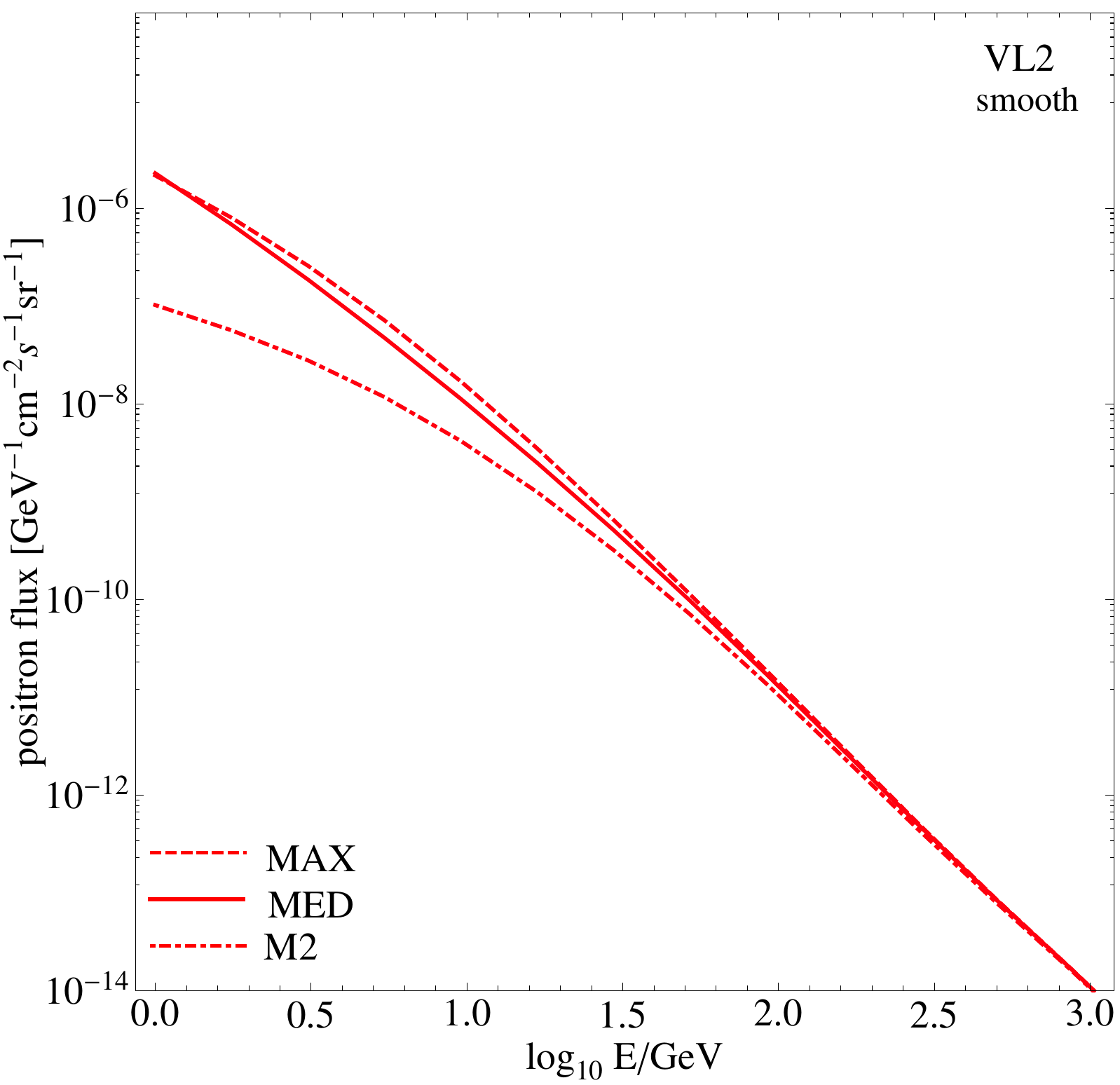}\\
 \includegraphics[width=7.5cm,height=7.5cm]{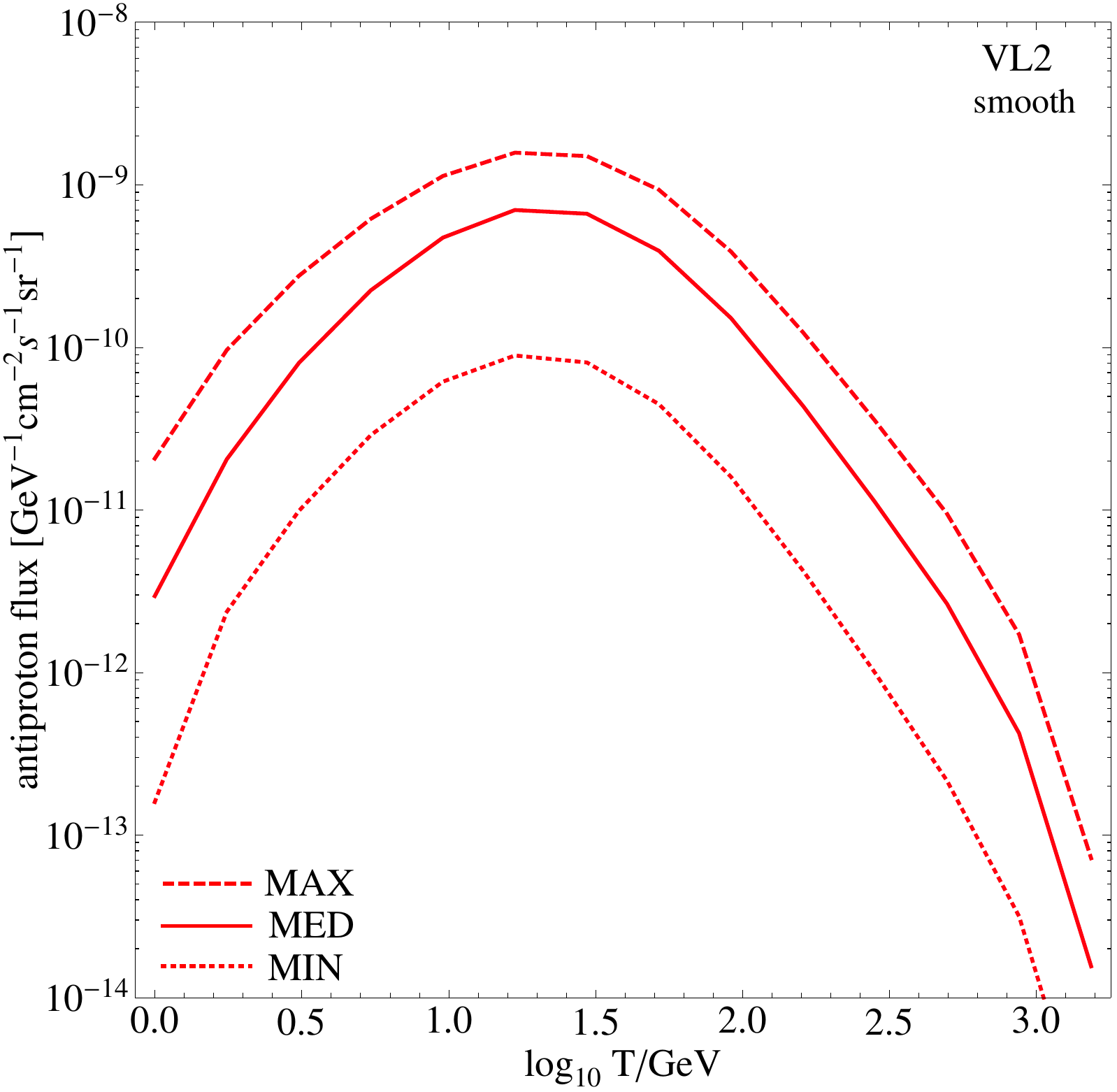}
 \caption{\fontsize{9}{9}\selectfont (Color online) The smooth quantity $(1-f_{\odot})^2 S(v_{\odot}) \phi_{sm}^0$ for the MDM3 candidate using the M2 (dot-dashed), MIN (dotted), MED (solid) and MAX (dashed) propagation models. For clarity, the clumpy component is omitted and just the results for Via Lactea II are plotted.}
 \label{figLS43prop}
\end{figure}

\par As an example we show in figure \ref{figLS43} the quantities $(1-f_{\odot})^2 S(v_{\odot}) \phi_{e^{+},sm}^0$ and $S_{max}\langle \phi_{e^{+},cl}^0 \rangle$ for the MDM3 candidate presented in table \ref{tab2}. Both Via Lactea II and Aquarius parameters are used and the MED propagation set is assumed. On the one hand, the smooth contribution with Aquarius is larger because its local DM density is higher than in Via Lactea II $-$ check table \ref{tab1}. On the other hand, the Via Lactea II simulation predicts (through extrapolation of $d^2N_{sh}/dMdV$) more low-mass clumps than Aquarius and thus the corresponding clumpy contribution is more significant. In fact, the ratio between the two clumpy fluxes is simple to understand if one computes the ratio
\begin{displaymath}
\frac{\left[\rho_{\odot}^2 \, N_{cl} \langle \xi \rangle_M \frac{dP_V}{dV}(r_{\odot})\right]_{\textrm{Aquarius}}}{\left[\rho_{\odot}^2 \, N_{cl} \langle \xi \rangle_M \frac{dP_V}{dV}(r_{\odot})\right]_{\textrm{Via Lactea II}}}\simeq 0.14 \, .
\end{displaymath}
Note that, as stated in section \ref{secnbody}, we are fixing $M_{min}=10^{-6}\textrm{ M}_{\odot}$; clumpy fluxes scale according to figure \ref{figNclxiM}.

\par Figure \ref{figLS43prop} shows the effect of varying the propagation parameters for the candidate considered above, for the Via Lactea II case without clumps.

\section{Antiprotons}\label{secpbar}

\par As in the previous section, we follow \cite{lavalle} to model the flux of antiprotons at the Earth. Differently from positrons, antiprotons do not lose much energy by
synchrotron or inverse Compton since $m_p\gg m_e$. Instead, they may be
swept away by galactic winds, assumed to be constant and perpendicular to the disk: $\vec{V}_c(\textbf{x})=sgn(z)V_c
\vec{e}_z$. Furthermore, annihilations $p\bar{p}$ are responsible for the
disappearance of primary antiprotons. These annihilations take place
essentially along the galactic plane where the interstellar medium is concentrated, and therefore the diffusion equation for antiprotons contains a term $-2h \delta_D(z) \Gamma_{ann}^{p\bar{p}} n_{\bar{p}}$ with
$\Gamma_{ann}^{p\bar{p}}=(n_H+4^{2/3}n_{He})\sigma_{ann}^{p\bar{p}}v_{\bar{p}}$,
$n_H\simeq 0.9 \textrm{ cm}^{-3}$, $n_{He}\simeq 0.1 \textrm{ cm}^{-3}$ \cite{Maurin06}
and 
\begin{align*}
& \sigma_{ann}^{p\bar{p}}(T_{\bar{p}})= \\
& \left\{
\begin{array}{ll}
661(1+0.0115T_{\bar{p}}^{-0.774} -0.948T_{\bar{p}}^{0.0151}) & T_{\bar{p}}<15.5\\
36 T_{\bar{p}}^{-0.5} & T_{\bar{p}}\geq 15.5
\end{array}
 \right. ,
\end{align*}
where $\sigma_{ann}^{p\bar{p}}$ is in mbarn, $T_{\bar{p}}=E_{\bar{p}}-m_p$ is the antiproton kinetic energy and
is in units of GeV in the above formula. The diffusion coefficient is
$K_{\bar{p}}(T_{\bar{p}})=K_0\beta_{\bar{p}}\left(\frac{p_{{\bar{p}}}}{\textrm{GeV}}\right)^{\delta}$
with
$\beta_{\bar{p}}=\left(1-\frac{m_p^2}{(T_{\bar{p}}+m_p)^2}\right)^{1/2}$
and $p_{\bar{p}}=(T_{\bar{p}}^2+2m_p T_{\bar{p}})^{1/2}$. All in all, the steady-state solution of the antiproton diffusion equation obeys 
\begin{align*}\label{diffpb}
-K_{\bar{p}}(T_{\bar{p}})\nabla^2 n_{\bar{p}}+ & \frac{\partial}{\partial z}\left(\textrm{sgn}(z) V_c n_{\bar{p}}\right)= \\
 & \qquad Q_{\bar{p}}(\textbf{x},T_{\bar{p}})-2h \delta_D(z) \Gamma_{ann}^{p\bar{p}}(T_{\bar{p}}) n_{\bar{p}} \, ,
\end{align*}
with $Q_{\bar{p}}$ of form analogous to \eqref{source}.

\par Once again, $\phi_{\bar{p}}(\textbf{x},T_{\bar{p}})=\frac{v_{\bar{p}}}{4\pi} n_{\bar{p}}(\textbf{x},T_{\bar{p}})$. Because we are interested in high-energy antiprotons we have neglected solar modulation and reacceleration effects. Similarly to the positron case, one has
\begin{equation}\label{smoothp}
\phi_{\bar{p},sm}^0(T)= \frac{v_{\bar{p}}}{4\pi} \frac{1}{2} \left(\frac{\rho_{\odot}}{m_{DM}}\right)^2 \, f_{inj}^{\bar{p}}(T) \, I_{sm}^{\bar{p}} (T) \, ,
\end{equation}
with
\begin{equation*} \label{Ismp}
I_{sm}^{\bar{p}} (T)= \int_{DZ}{d^3 \textbf{x} \,  \left(\frac{\rho_{sm}(\textbf{x})}{\rho_{\odot}}\right)^2 \, G_{\odot}^{\bar{p}}(\textbf{x},T,L,K_0,\delta,V_c)} \, ,
\end{equation*}
and
\begin{equation}\label{clumpyp}
\langle \phi_{\bar{p},cl}^0 \rangle (T)= \frac{v_{\bar{p}}}{4\pi}  \frac{1}{2} \left(\frac{\rho_{\odot}}{m_{DM}}\right)^2 N_{cl} \, \langle \xi \rangle_M \, f_{inj}^{\bar{p}}(T) \, \langle G_{\odot}^{\bar{p}} \rangle_V (T) \, ,
\end{equation}
where
\begin{equation*} \label{Iclp}
\langle G_{\odot}^{\bar{p}} \rangle_V (T)= \int_{DZ}{d^3 \textbf{x} \, G_{\odot}^{\bar{p}}(\textbf{x},T,L,K_0,\delta,V_c)} \, \frac{dP_V}{dV}(\textbf{x}) \, .
\end{equation*}
The Green function for antiprotons $G_{\odot}^{\bar{p}}$ is given in \cite{lavalle}. At last, the total antiproton flux is
\begin{equation}
\phi_{\bar{p}}(T)=(1-f_{\odot})^2 \, S(v_{\odot}) \, \phi_{\bar{p},sm}^0 (T) \, + \, S_{max} \, \langle \phi_{\bar{p},cl}^0 \rangle (T) \, .
\end{equation}

\par Figures \ref{figLS43} and \ref{figLS43prop} show the antiproton fluxes for the minimal dark matter triplet considered at the end of the previous section.

\begin{figure}
 \centering
 \includegraphics[width=7.5cm,height=7.5cm]{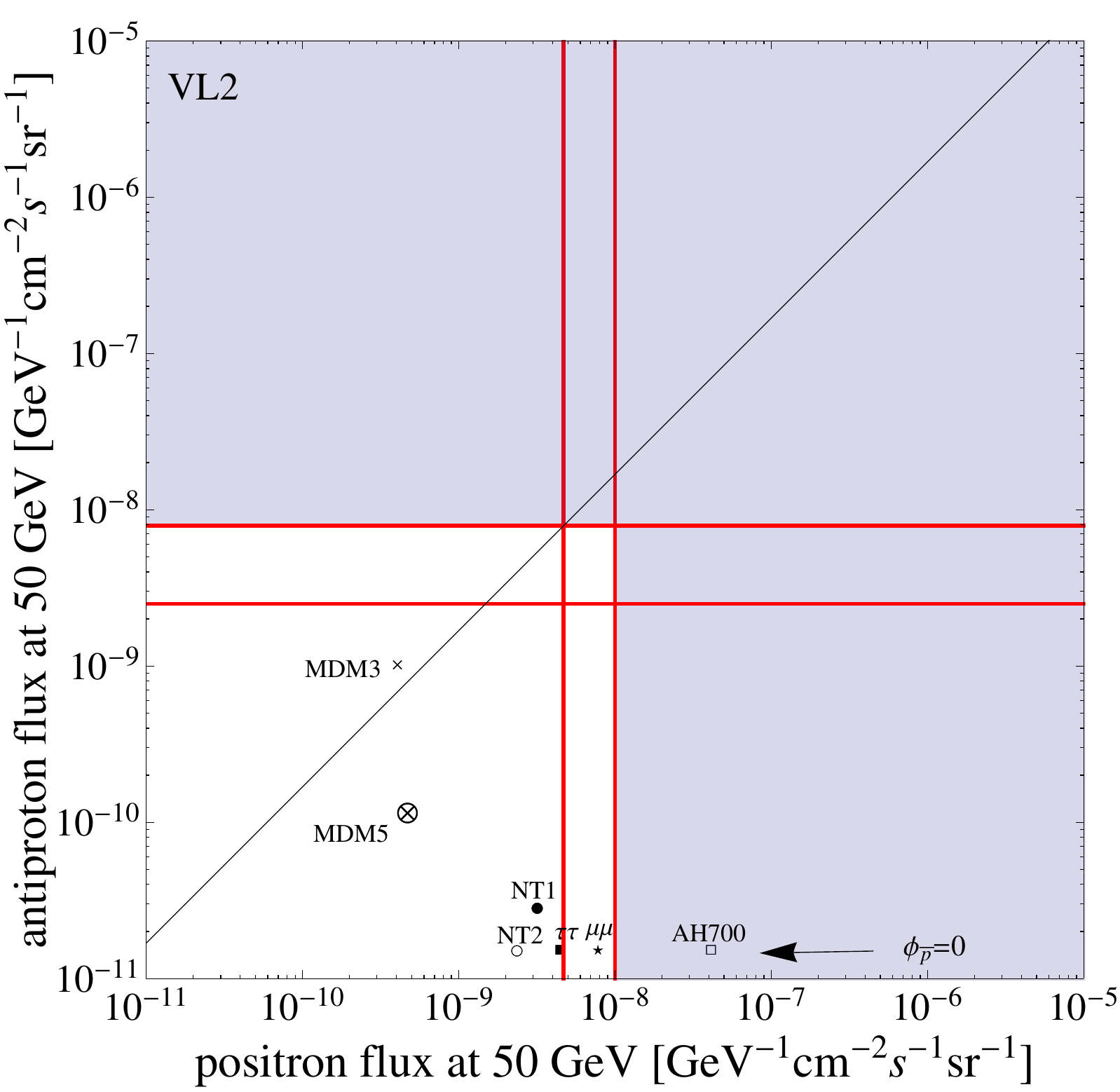}\\
\includegraphics[width=7.5cm,height=7.5cm]{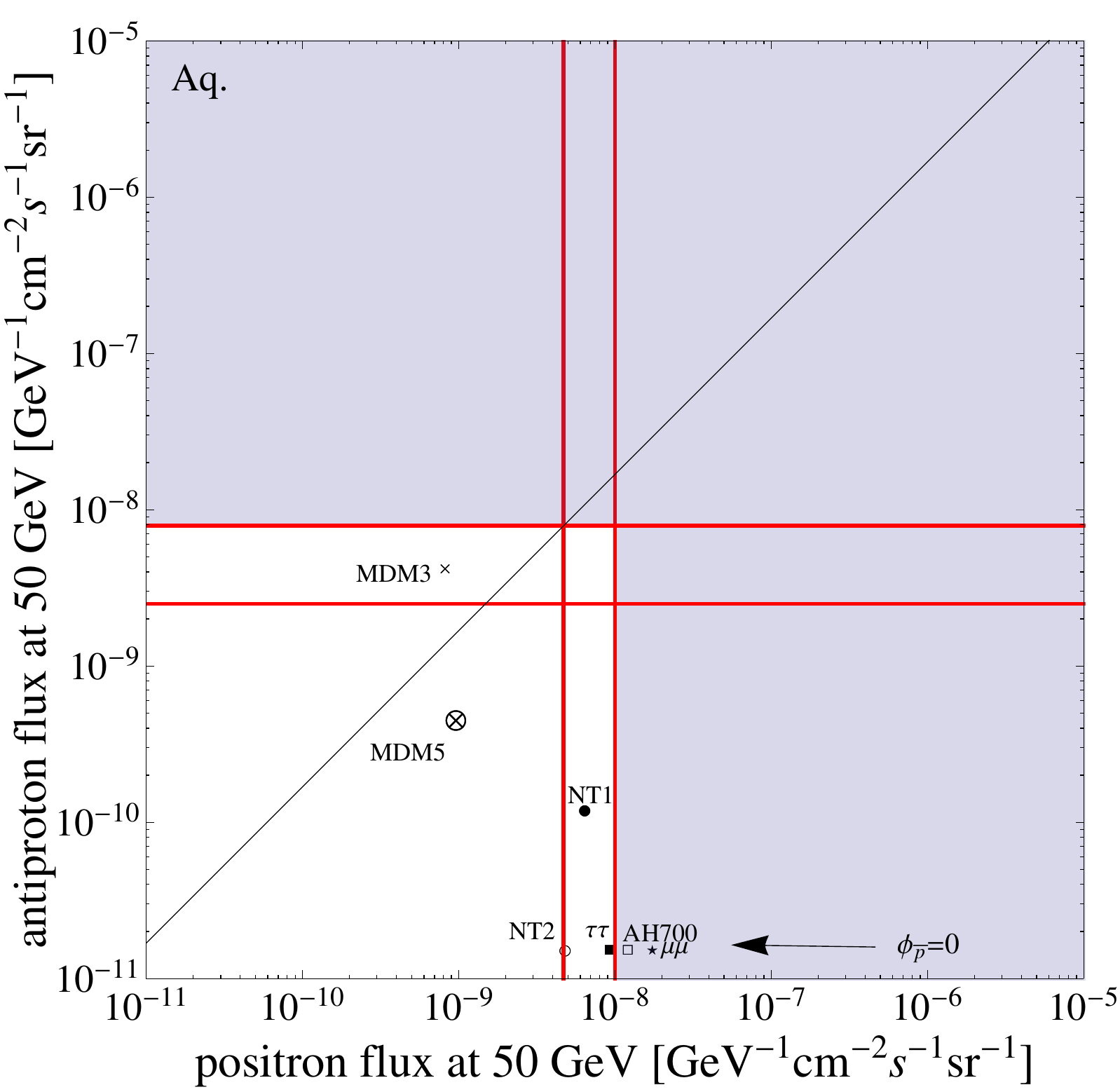}
 \caption{\fontsize{9}{9}\selectfont (Color online) Total positron and antiproton fluxes at 50 GeV for the models in table \ref{tab2} and for both Via Lactea II and Aquarius parameters. The MED propagation is used and $M_{min}=10^{-6} \textrm{ M}_{\odot}$. The solid lines indicate the fluxes deduced from PAMELA data assuming electron and proton fluxes at 50 GeV $-$ see equations \eqref{pamelae}, \eqref{pamelap} $-$ and the corresponding excluded regions are shown as shaded. Candidates lying above the diagonal line cannot be rescaled to explain PAMELA positron excess without overproducing antiprotons.}\label{fige+pbar}
\end{figure}

\vspace{0.5cm}

\par For clarity in visualising the $e^+$ and $\bar{p}$ yields from the candidates in table \ref{tab2}, we choose to plot fluxes at 50 GeV which is well inside the energy range where PAMELA detected the positron excess and collected antiprotons. Moreover, both positrons and antiprotons of such energy are not significantly affected by solar modulation or reacceleration effects. At 50 GeV the PAMELA data indicates \cite{pamela,pamelapbar} $\phi_{e^+}/(\phi_{e^-}+\phi_{e^+})\sim 0.1$ and $\phi_{\bar{p}}/\phi_{p}\sim 0.7-2.2 \cdot 10^{-4}$ that we translate using figures 9 and 10 in \cite{Delahaye08} and equation 1 in \cite{Donato08} into the fluxes:
\begin{equation}\label{pamelae}
\tilde{\phi}_{e^+}(50 \textrm{ GeV})=\left\{ 
\begin{array}{ll}
1.0 \cdot 10^{-8}  &  \textrm{for hard } e^-\\
4.7 \cdot 10^{-9} &  \textrm{for soft } e^-\\
\end{array} \right. \, , \textrm{ and}
\end{equation}
\begin{equation}\label{pamelap}
\tilde{\phi}_{\bar{p}}(50 \textrm{ GeV})=\left\{ 
\begin{array}{ll}
7.9 \cdot 10^{-9} &  \textrm{for } \phi_{\bar{p}}/\phi_{p}=2.2 \cdot 10^{-4} \\
2.5 \cdot 10^{-9} &  \textrm{for } \phi_{\bar{p}}/\phi_{p}=0.7 \cdot 10^{-4}\\
\end{array} \right. \, ,
\end{equation}
in units of $\textrm{ GeV}^{-1}\textrm{cm}^{-2}\textrm{s}^{-1}\textrm{sr}^{-1}$.

\par Our aim in the present work is not to perform a fitting procedure or a likelihood analysis to PAMELA (or ATIC) data, but rather to investigate which DM particles are able to produce positron fluxes near the above-stated values. Figure \ref{fige+pbar} shows the total positron and antiproton fluxes at 50 GeV for the models in table \ref{tab2} and assuming the MED propagation configuration and $M_{min}=10^{-6}\textrm{ M}_{\odot}$. We see immediately that some models violate the positron and/or antiproton data. However, all candidates may present $(\sigma_{ann} v)_0$ a few times more or less than presented in table \ref{tab2}, which accounts for a scaling along diagonals in figure \ref{fige+pbar} since a change of $(\sigma_{ann} v)_0$ modifies equally the positron and the antiproton fluxes. Conservatively, we are interested in knowing which points in figure \ref{fige+pbar} may be rescaled to touch the left vertical line without being above the upper horizontal line. In other words, we wish to pin down the particles that can meet the PAMELA positron excess without violating antiproton bounds. Candidates lying above the diagonal line in figure \ref{fige+pbar} cannot. For the DM distribution suggested by Via Lactea II the model labelled MDM3 is disfavoured. The situation for Aquarius is similar, but more constraining. Lastly, leptophilic candidates produce no antiprotons and automatically pass the $\bar{p}$ test; they are plotted in figure \ref{fige+pbar} for completeness and with an artificial $\phi_{\bar{p}}$. The exception is NT1 model that features a non-zero BR($s\to b\bar{b}$), even though the corresponding antiproton flux is rather low.

\par As mentioned above, figure \ref{fige+pbar} refers to the MED propagation parameters. From figure \ref{figLS43prop} one sees that at 50 GeV the positron flux is not very sensitive to the propagation parameters, while the antiproton flux at 50 GeV may be roughly one order of magnitude above or below the flux computed with MED. Since the antiproton bound will turn out to be the less constraining one, we can safely stick to the mean propagation in presenting our main results.

\par Next we proceed with constraints coming from $\gamma$-ray observations.

\section{Gamma-rays}

The expected $\gamma$-ray flux from DM annihilation is proportional to the line of sight integral (LOS) of the DM density square $\rho_\chi^2$ $-$ which tells us how many annihilations we have in the cone of view defined by the experimental angular resolution $-$ and to the annihilation cross-section $-$  which gives us the yields of photons obtained in one annihilation:
\begin{equation}\label{flussodef}
\frac{d \Phi_\gamma}{dE_\gamma}(M,E_\gamma, \psi, \theta) =
 \frac{1}{4 \pi} \frac{\sigma_{ann} v }{2 m^2_\chi} \cdot 
\sum_{f} \frac{d N^f_\gamma}{d E_\gamma} B_f  \int_{V}  \frac{\rho_\chi^2(M,R)}{d^2} dV \, .
\end{equation}

For each halo along our line of sight, $M$ is the mass and $d$ the distance from the observer. 
$d N^f_\gamma / dE_\gamma$ is the differential spectrum per annihilation of prompt photons \cite{FPS04,bergstrom} relative to the final state $f$, with branching ratio $B_f$. 
The volume integral refers to the line of sight and is defined by the angular resolution of the instrument $\theta$ and by the direction $\psi$ of observation. 
Inside each halo, $R$ is the distance from the centre. $\rho_\chi(M, R, c(M,R))$ is the DM density profile with concentration parameter $c(M,r)$.

When including the Sommerfeld enhancement of the local annihilation cross section, the term $\sigma_{ann} v $  in eq. \eqref{flussodef}  is replaced by the velocity-dependent expression $(\sigma_{ann} v)_0 S(R,M)$.
The enhancement $S$ depends on the halo mass fixing the average velocity dispersion, and from the radial coordinate inside the halo, which takes into account the features of the velocity dispersion curve that has lower values closer to the centre of the galaxy hosted by the DM halo $-$ see e.g. \cite{zentner}.

The enhancement will be included in the line of sight integral, which transforms into
\begin{equation}
\Phi_S^{LOS} =  \int \frac{S(M,R) \rho_\chi^2(M,c,R) }{d^2} dV \, .
\label{flussodefS}
\end{equation}

In the next paragraphs we will compute the prompt $\gamma$-rays coming from DM annihilation in 1) the smooth halo of our galaxy, 2) the substructures of our galaxy and 3) the extragalactic halos and subhalos.

\subsection{Annihilations in the smooth MW halo}
The astrophysical contribution to the prompt $\gamma$-ray emission from the smooth component of a DM halo can be rewritten as the volume integral
\begin{equation}
\Phi_S(M, d, \psi, \Delta \Omega) \propto \int \int \int_{V}  
d \phi d \theta d\lambda  \left [ \frac{S(M,R) \rho_{\chi}^2(M,R)} {\lambda^{2}}\right] \, \nonumber
\end{equation}
where  $\lambda$ is the line-of-sight coordinate,  $\Delta \Omega$ the solid angle corresponding to the angular resolution $\theta$ of the instrument, and $\psi$ the angle of view from the GC. Both in the case of ACTs and of Fermi (E $>$ 1 GeV) we will use $\theta = 0.1^\circ$ and  $\Delta \Omega \sim 10^{-5} \sr$. 

\begin{figure}
 \centering
 \includegraphics[width=7.5cm,height=7.5cm]{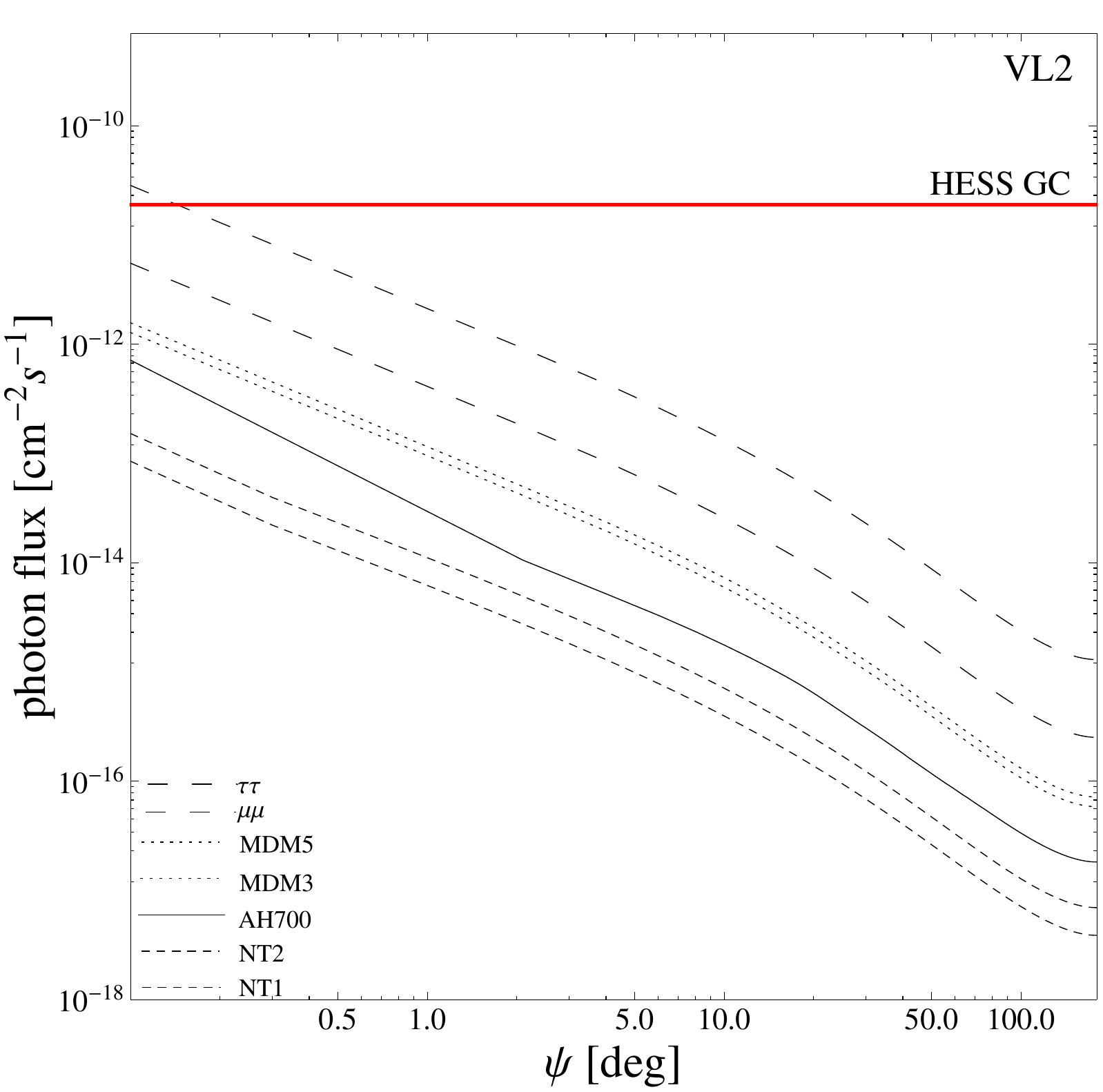}\\
 \includegraphics[width=7.5cm,height=7.5cm]{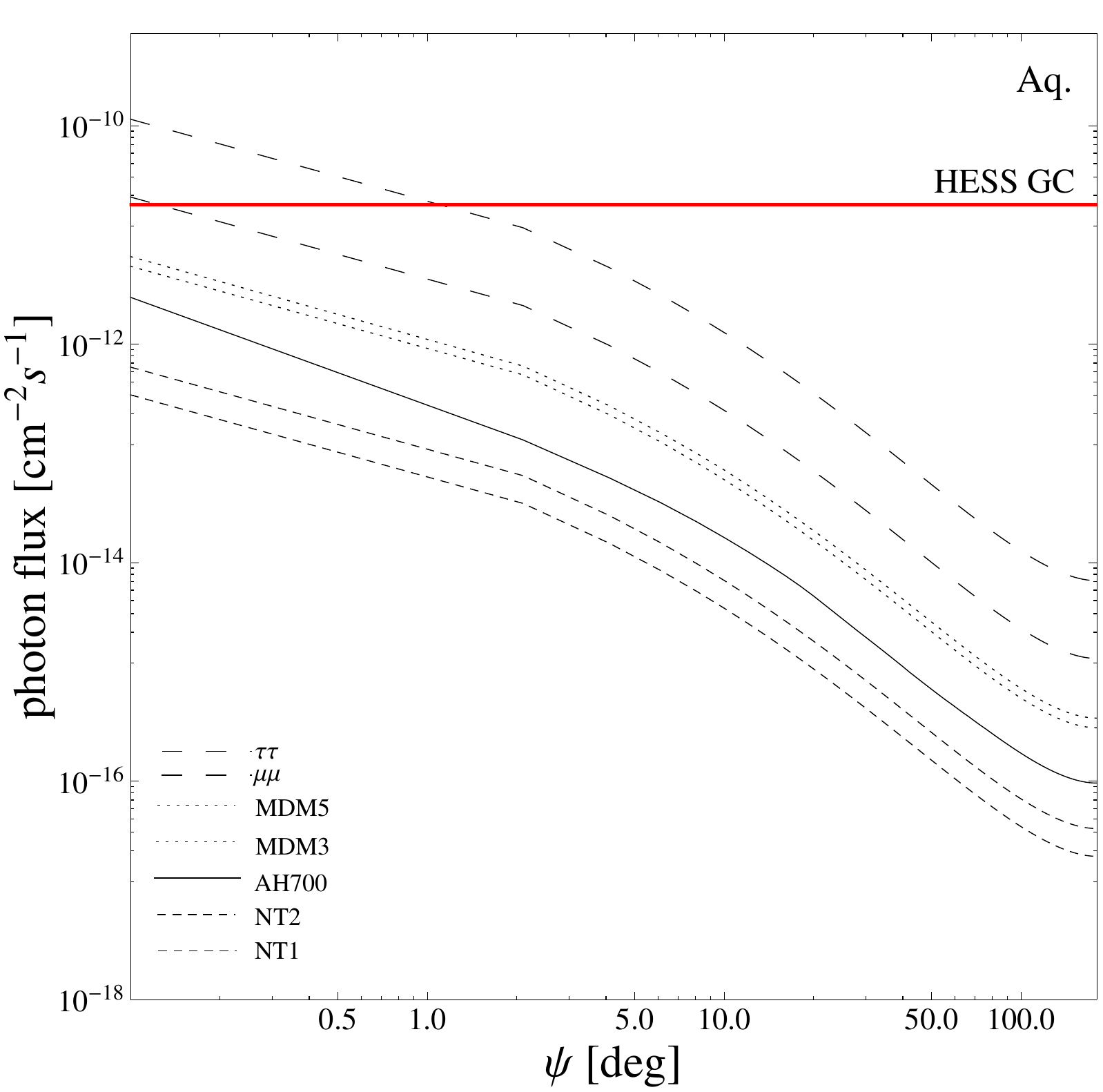}\\
\vspace{0.2cm}
\fontsize{9}{9}\selectfont
\begin{tabular}{l|c|c}
\hline
\hline
 & \multicolumn{2}{c}{$\phi_{MW} (\psi=0^{\circ})$ [cm$^{-2}$s$^{-1}$]} \\
\cline{2-3}
label & Via Lactea II & Aquarius \\ 
\hline
AH700 & $1.15 \cdot 10^{-12}$ & $3.38 \cdot 10^{-12}$ \\
NT1 & $1.33 \cdot 10^{-13}$ & $4.16 \cdot 10^{-13}$ \\
NT2 & $2.39 \cdot 10^{-13}$ & $7.47 \cdot 10^{-13}$ \\
$\mu\mu$ & $8.67 \cdot 10^{-12}$ & $2.70\cdot 10^{-11}$\\
$\tau\tau$ & $4.47 \cdot 10^{-11}$ & $1.39 \cdot 10^{-10}$ \\
\hline
MDM3 & $2.01 \cdot 10^{-12}$ & $6.25 \cdot 10^{-12}$ \\
MDM5 & $2.47 \cdot 10^{-12}$ & $7.69 \cdot 10^{-12}$ \\
\hline
HESS GC & \multicolumn{2}{c}{$1.89\cdot 10^{-11}$} \\
\hline
\end{tabular}
\caption{\fontsize{9}{9}\selectfont (Color online) The $\gamma$-ray flux above 160 GeV as a function of the angle $\psi$ with respect to the GC. The legend in the figures is ordered according to the values of the curves at $\psi=0.1^{\circ}$. The HESS measurement towards the GC is shown by the horizontal thick line and the table shows the fluxes for $\psi=0^{\circ}$.}\label{figGamma1}
\end{figure}

The HESS telescope has observed the GC source in 2003 and 2004, measuring an integrated flux above 160 GeV of $\Phi (> 160 \GeV) = 1.89 \times 10^{-11} \gamma \cm^{-2} \sec^{-1}$ \cite{hessgc}. In figure \ref{figGamma1} we show the result of our computation for the particle physics models of table \ref{tab2}, in the cases where the Milky Way halo is described by either the VL2 or the Aquarius models.
Also shown is a table with the expected flux above 160 GeV at the GC, for direct comparison with the HESS limit. Considering $(\sigma_{ann} v)_0 = 3 \times 10^{-26} \cm^{3} \sec^{-1}$, the $\tau\tau$ candidate is ruled out.

\subsection{Annihilations in the subhalos of the MW}

We now populate the halo of our galaxy with subhalos with masses as small as $10^{-6} M_{\odot}$.

The LOS contribution of such a population of substructures to the annihilation signal can be 
written as
$$ 
\Phi_S^{\rm sub}(M_{h}, d, \psi, \Delta \Omega) \propto \int_{M_ {sub}} d M_ {sub} \int_c d c \int \int_{\Delta \Omega} 
d \theta d \phi 
$$
\begin{equation}
\int_{\lambda}  d\lambda [ \rho_{sh}(M_{h},M_{sub},R) P(c)  \Phi_{S}^{halo}]
\label{smoothphicosmo}
\end{equation}
where the contribution from each subhalo $\Phi_{S}^{halo} (M,S(M,R),c(M,R),d,\psi, \Delta \Omega$) is convolved with its distribution function inside the galaxy $\rho_{sh}$. Here R is the radial coordinate with respect to the centre of the host galaxy.
$P(c)$ is the lognormal distribution of the concentration parameter with dispersion 
$\sigma_c$ = 0.14 \cite{Bullock01} and mean value $ \bar{c}$:
\begin{equation}
P(\bar{c},c) = \frac{1}{\sqrt{2 \pi} \sigma_c c} \, 
e^{- \left ( \frac{\ln(c)-\ln(\bar{c})} {\sqrt{2} \sigma_c} \right )^2}.
\end{equation}
The satellites inside a galaxy suffer from external tidal stripping due to the interaction with the gravitational potential of the galaxy itself.
To account for gravitational tides, we follow \cite{Hayashi} and assume that all the mass beyond the subhalo tidal radius is lost in a single orbit without affecting its central density profile. The tidal radius is defined as the distance from the subhalo centre at which the tidal forces of the host potential equal the self-gravity of the subhalo. In the Roche limit, it is expressed as:
\begin{equation}
r_{tid}(r)= \left (\frac{M_{sub}}{2 M_{h}(<r)} \right)^{1/3} r 
\end{equation}
where r is the distance from the halo centre, $M_{sub}$ the subhalo mass and $M_{h}(<r)$ the host halo mass enclosed in a sphere of radius r. The integral along the line of sight will be different from zero only in the interval [$d - r_{tid}, d + r_{tid}]$.

\begin{figure}
 \centering
 \includegraphics[width=7.5cm,height=7.5cm]{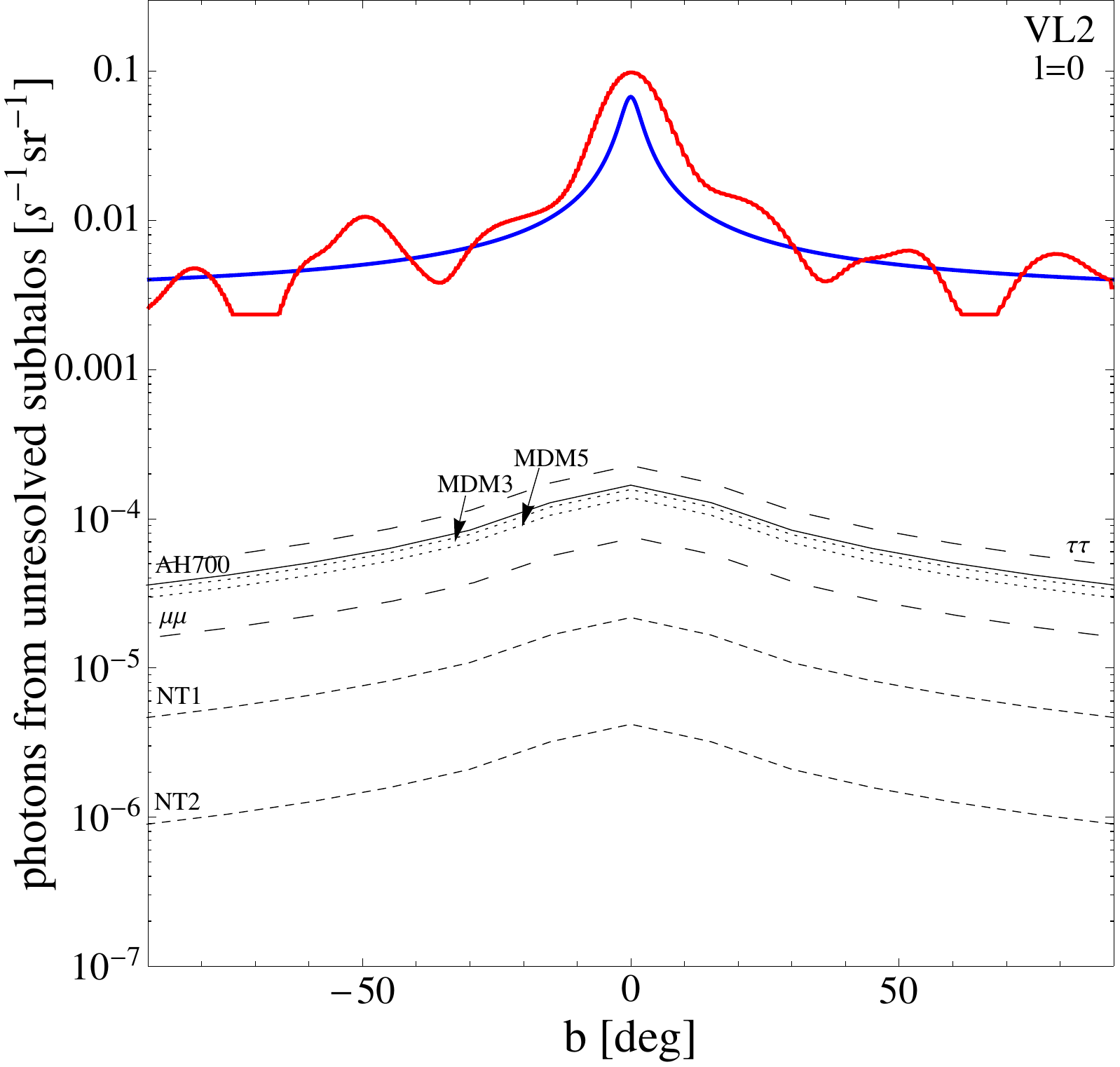}\\
 \includegraphics[width=7.5cm,height=7.5cm]{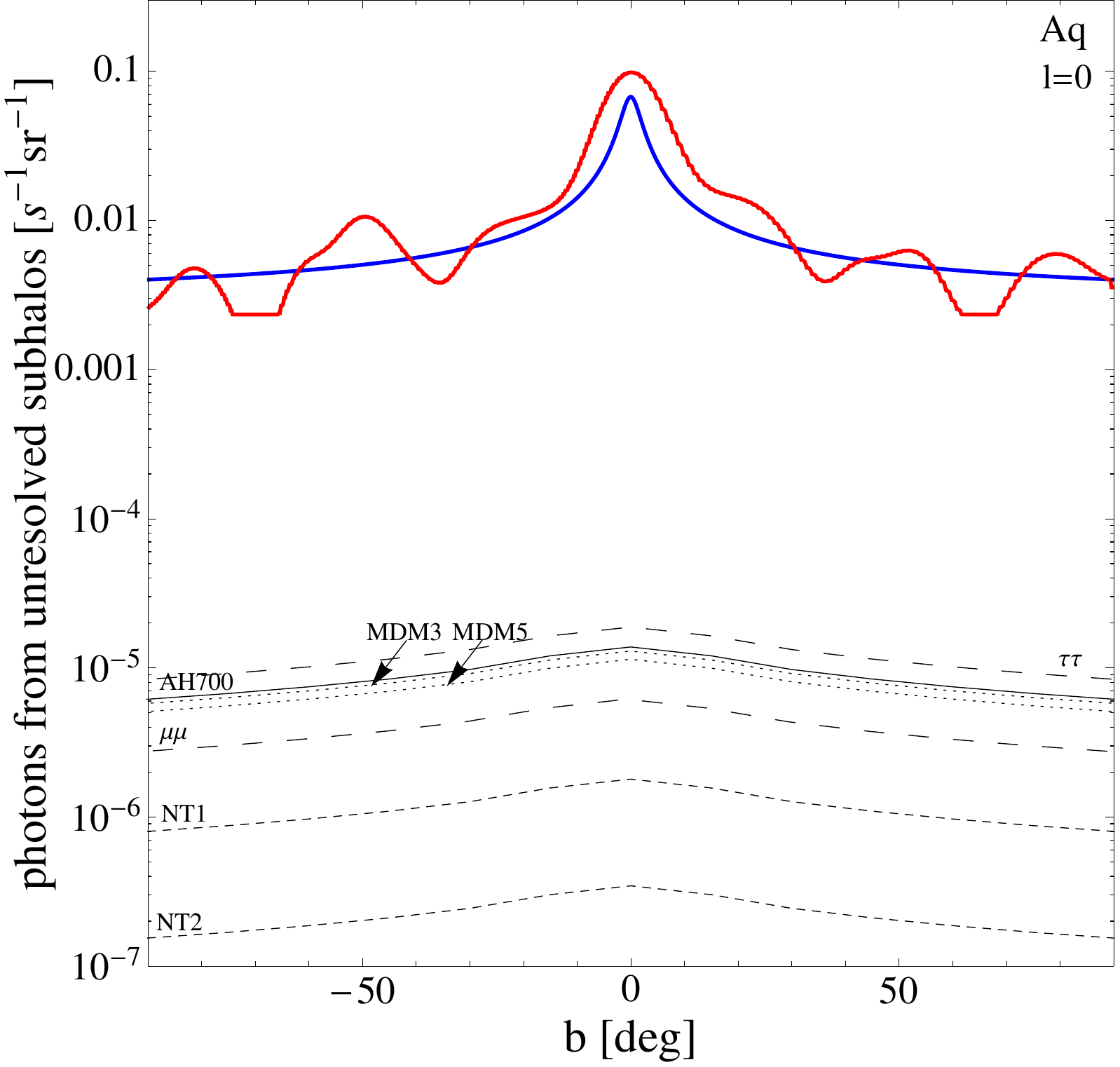}
 \caption{\fontsize{9}{9}\selectfont (Color online) Number of photons above 3 GeV from unresolved subhalos at a galactic longitude $l=0^{\circ}$ for Via Lactea II and Aquarius parameters. Again, we fix $M_{min}=10^{-6} \textrm{ M}_{\odot}$. Shown in thick solid are the EGRET map for diffuse background as well as the smooth approximation of reference \cite{bergstrom98}, both scaled to Fermi results.}\label{figGamma2}
\end{figure}

The estimate of the $\gamma$-ray flux from substructures is obtained through the numerical integration  of Eq. \eqref{smoothphicosmo}.
The result of our computations is shown in figure \ref{figGamma2}, where the expected flux from the subhalos is compared with the EGRET diffuse emission (galactic + extragalactic) background.
Although the EGRET flux has been rescaled by $\sim$15\% in the light of the new measurement from the Fermi telescope which does not confirm the GeV bump, the diffuse emission measurements turn out to be less constraining than the GC one.

\subsection{Annihilations in the extragalactic halos and subhalos}

Any direction of observation will also receive a contribution from all the halos and subhalos
at all redshift along the line of sight. 
We adapt the formalism of Ref.~\cite{Ullio:2002pj} to estimate their contribution to the prompt $\gamma$-ray
flux in case of Sommerfeld enhancement. 
We compute the extragalactic contribution at E = 10 GeV, and compare it with the EGRET extragalactic 
background at the same energy, rescaled for a 40\% due to the contribution of those unresolved blazars which have been now observed and included in the Fermi catalogue, namely $\phi^{extragal}_{EGRET}(E=10 \GeV) = 6 \times 10^{-9} \cm^{-2} \sec^{-1}$.

The prompt photon flux from DM annihilation in the extragalactic host halos can be written as:
\begin{eqnarray}
& &  \frac{d\Phi}{dE_0d\Omega} (E_0) \equiv
\frac{d\mbox{\sffamily{N}}_\gamma}{dE_0 dt_0 dA d\Omega} 
\label{eqn:dPhi_dE_extra}
\end{eqnarray}
where $ E_0 $ and $ dt_0 $ are the energy and the time interval
over which the photons are detected on earth.
$d\mbox{\sffamily{N}}_\gamma $ is the number of $\gamma$-ray photons produced in the infinitesimal
volume $dV$ at redshift $z$ ($dV \propto d\Omega (1+z)^{-3}$), in a 
time interval $ dt $ ($dt = dt_0 (1+z)^{-1}$) with an energy between $ E $ and $E+dE$ ($E=E_0 (1+z))$ and collected by a detector with effective area $dA$. 

$d\mbox{\sffamily{N}}_\gamma $  is obtained integrating the single halo contribution to the photon flux
$d\mathcal{N}_\gamma/dE$ over the halo mass function $\frac{dn}{dM}(M,z)$: 
\begin{eqnarray}
d\mbox{\sffamily{N}}_\gamma & \propto & e^{-\tau(z,E_0)} \left[ (1+z)^3 \int dM 
\frac{dn}{dM}(M,z) \right. \\ \nonumber 
& & \left. \frac{d\mathcal{N}_\gamma}{dE}(E,M,z) \right].
\label{eqn:dN}
\end{eqnarray}
Following the Press-Schechter formalism \cite{Press:1973iz}, the halo mass function is given by
\begin{equation}
\frac{dn}{dM}(M,z)=\frac{\rho_{\mbox{\tiny{cr}}}\Omega_{0,m}}{M^2}
\nu f(\nu) \frac{d\log\nu}{d\log M},
\label{eqn:dndM}
\end{equation}
where $ \rho_{\mbox{\tiny{cr}}} $ is the critical density, $\Omega_{0,m}$ is 
the mass density parameter, $\nu=\frac{\delta_{sc}(z)}{\sigma(M)}$, 
$ \sigma(M) $ is the rms density fluctuation on the mass scale $M$
and $\delta_{sc}$ represents the critical density for spherical collapse \cite{Ullio:2002pj,Ahn:2007ty,Eisenstein:1997jh}. \\

$ d\mathcal{N}_\gamma/dE $ is the number of photons with energy 
between $ E $ and $E+dE $ produced in a halo of mass $ M $ at redshift $ z $, and can be written as
\begin{eqnarray}
\frac{d\mathcal{N}_\gamma}{dE}(E,M,z) & = & \frac{\sigma_{ann} v}{2} 
\frac{dN_\gamma(E)}{E} \frac{M}{m_\chi^2} 
\frac{\Delta_{vir} \rho_{\rm cr} \Omega_m(z)}{3} \\ \nonumber 
& & \frac{c^3(M,z)}{I_1(x_{\mbox{\tiny{min}}},c(M,z))^2} 
I_2(x_{\mbox{\tiny{min}}},c(M,z)). 
\label{eqn:dmathcalNdE}
\end{eqnarray}
In the previous expression, the virial overdensity is \cite{Ullio:2002pj}:
\begin{equation}
\Delta_{vir}(z)=\frac{18 \pi^2 + 82(\Omega_m(z)-1) -39
(\Omega_m(z)-1)^2}{\Omega_m(z)}
\end{equation}
and 
the integrals $ I_1 $ and $ I_2 $ enter the LOS integral
\begin{equation}
I_n(x_{\mbox{\tiny{min}}},x_{\mbox{\tiny{max}}})=\int g^n x^2 dx,
\end{equation}
where  $ g(x)=x^{-1}(1+x)^{-2} $ in the case of an NFW profile and 
$ g(x)=e^{-2 (x^\alpha -1)/\alpha} $ in the Einasto case. \\

Finally, the absorption coefficient $e^{-\tau(z,E_0)} $, $\tau(z,E_0)=z/[3.3(E_0/10 \mbox{ GeV})^{-0.8}] $ accounts for pair production due to the interaction of the $\gamma$-ray photons with the 
extra-galactic background light in the optical and infrared bands \cite{Bergstrom:2001jj}. \\

With a little algebra, we get to the final expression for the extragalactic DM $\gamma$-ray flux:
\begin{eqnarray}
\label{eqn:aveflux}
 \frac{d\Phi}{dE_0d\Omega}(E_0) = 
\frac{\sigma_{ann} v}{8\pi}\frac{c}{H_0}
\frac{\rho_{\mbox{\tiny{cr}}}^2 \Omega_{0,m}^2}{m_\chi^2} \times  \ \ \ \ \ \ \ \ \ \  \\
 \int dz (1+z)^3 \frac{\Delta^2(z)}{h(z)} \frac{dN_\gamma(E_0(1+z))}{dE}  e^{-\tau(z,E_0)}, \nonumber
\end{eqnarray}
with
\begin{equation}
\Delta^2(z)=\int dM \frac{\nu(z,M)f(\nu(z,M))}{M\sigma(M)} \left|
\frac{d\sigma}{dM} \right| M \Delta_M^2(z,M)
\label{eqn:Delta2}
\end{equation}
and
\begin{equation}
\Delta_M^2(z,M)=\int dc^\prime P(c(M,z),c^\prime)
\frac{\Delta_{vir}}{3} \frac{I_2(x_{\mbox{\tiny{min}}},c^\prime)}
{I^2_1(x_{\mbox{\tiny{min}}},c^\prime)} (c^{\prime})^3 dc^\prime.
\label{eqn:Delta2M}
\end{equation}
$ \Delta^2(z) $ is a quantity which describes the boost to the isotropic $\gamma$-ray background
due to the existence of virialized DM halos. \\

To account for the presence of substructures inside the extragalactic host halos, we have to apply the substitution
\begin{equation}
\label{eqn:substructures}
M \Delta_M^2(M) \rightarrow  (1-f(M))^2 M \Delta_M^2(M)+ \Delta_{M,sub}^2(M) 
\end{equation}
where $f(M)$ is the fraction of mass in virialized  substructures within a host halo of mass $M$, and can be expressed as 
\begin{equation}
f(M)_{VL2} = 1.45 \times 10^{-2} [log(M) +9.21] \, ,\nonumber
\end{equation}
\vspace{-0.5cm}
\begin{equation}
f(M)_{Aq} = 0.285 M^{-0.1} [0.63 M^{0.1}  - 0.25] \, . \nonumber
\end{equation}

More precisely, taking into account the effect of the radial dependence of the concentration parameter
of subhalos, we will have 
\begin{eqnarray}
\Delta_{M,sub}^2(z,M) & = & \int_{M_{sub}} dM_{sub} \int dc^\prime P(c(M,z),c^\prime) \\ \nonumber
&&  \int_0^{R_{vir}(M)}
4 \pi r^2 dr \rho_{sh}(M,r) \\ \nonumber 
&& \frac{\Delta_{vir}}{3} \frac{I_2(x_{\mbox{\tiny{min}}},c^\prime(M,r))}
{I^2_1(x_{\mbox{\tiny{min}}},c^\prime(M,r))} (c^{\prime}(M,r))^3 dc^\prime.
\label{eqn:Delta2Msub}
\end{eqnarray}

We obtain our results by solving numerically Eq.~\eqref{eqn:aveflux} both for host halos and for subhalos, and combining them through Eq.~\eqref{eqn:substructures}. The extragalactic bounds turn out to be the less constraining. Note that our bounds refer to prompt $\gamma$-rays only. However, DM annihilations in extragalactic halos and subhalos can also give rise to a $\gamma$-ray flux by producing high-energy electrons and positrons that up-scatter CMB photons. This inverse Compton scattering contribution is particularly relevant for leptophilic models and has been computed in \cite{ProfumoJeltema,BelikovHooper2} where it has been shown that COMPTEL and EGRET extragalactic observations place interesting limits on DM annihilation cross-sections. These constraints are competitive with the ones derived here with prompt $\gamma$-rays from the Galactic Centre. Another effect of the presence of inverse Compton photons is the ionisation of the baryonic gas after recombination and thus the decrease of the CMB optical depth $-$ see Refs.~\cite{BelikovHooper1,CIP09}.

\section{Synchrotron radiation}\label{secsync}
\par Synchrotron emission arises from relativistic electrons and positrons propagating in the galactic magnetic field. Since all annihilation channels usually considered produce high-energy electrons and positrons, a DM-induced synchrotron signal is expected from regions of the galaxy where a substantial magnetic field is active and the dark matter density is significant $-$ see e.g.~\cite{bi,bertone1,bergstrom}. Let us focus on a region towards the galactic centre, small enough so that diffusion does not play an important role and where the galactic magnetic field is strong enough to neglect electron (and positron) energy losses other than synchrotron emission. Assuming further that advection is negligible as in \cite{bertone1,bergstrom}, equation \eqref{diffe+} is easily solved in steady-state conditions for positrons and electrons:
\begin{equation}
n_{e^{\pm}} (\textbf{x},E)=\frac{\langle \sigma_{ann} v \rangle}{2 m_{DM}^2} \rho_{DM}^2 (\textbf{x}) \frac{N_{e^{\pm}}(>E)}{b_{syn}(\textbf{x},E)} \, ,
\label{ne+e-}
\end{equation}
where $N_{e^{\pm}}(>E)=\int_{E}^{m_{DM}}{dE' \, \frac{dN_{e^{\pm}}}{dE_{e^{\pm}}}(E')}$ is the number of electrons plus positrons per DM annihilation above a given energy $E$, $b_{syn}(\textbf{x},E)\simeq e^4 B^2(\textbf{x}) E^2/(9\pi m_e^4)$ and $B$ is the galactic magnetic field. Now, each electron (or positron) emits synchrotron radiation according to the power spectrum \cite{bertone1,bergstrom}
\begin{equation}
\nu \frac{d\tilde{W}_{syn}}{d\nu}(\textbf{x},E)=\nu \, \frac{\sqrt{3}}{6\pi} \frac{e^3 B(\textbf{x})}{m_e} \cdot \frac{8\pi}{9\sqrt{3}}\delta\left(\frac{\nu}{\nu_{syn}}-0.29\right) \, ,
\label{nudwdnu1}
\end{equation}
being $\nu_{syn}(\textbf{x},E)=3eB(\textbf{x})E^2/(4\pi m_{e}^3)$ the synchrotron frequency. The next step is to convolute equations \eqref{ne+e-} and \eqref{nudwdnu1} in the volume of observation $V_{obs}$, in order to give the total synchrotron power emitted by the distribution of DM-induced electrons and positrons:
\begin{align}\label{nudwdnu}
& \nu \frac{d W_{syn}}{d\nu}=\int_{V_{obs}}{dV  \int_{m_e}^{m_{DM}} {dE \, n_{e^{\pm}} (\textbf{x},E) \, \nu \frac{d\tilde{W}_{syn}}{d\nu}(\textbf{x},E)  } } \nonumber \\
& = \frac{\langle \sigma_{ann} v \rangle}{2 m_{DM}^2} \int_{V_{obs}}{dV \, \rho_{DM}^2 (\textbf{x}) \, E_{p}(\textbf{x},\nu) \, \frac{N_{e^{\pm}}(>E_p)}{2}  } \, ,
\end{align} 
where $E_p(\textbf{x},\nu)=\sqrt{\frac{4\pi m_e^3 \nu}{3\cdot 0.29 e B(\textbf{x})}}$.

\par It was pointed out in \cite{bertone1} that low-frequency radio constraints do not depend much on the magnetic field profile adopted. Following that work, we choose to implement a constant $B=7.2$ mG inside the accretion region $r\leq R_{acc}=0.04$ pc, $B\propto r^{-2}$ for $R_{acc}<r<84.5 R_{acc}$ and $B=\mu$G for $r\geq84.5 R_{acc}$.

\par In order to place constraints on $\frac{\sigma_{ann} v}{m_{DM}^2}$, we consider the three configurations studied in \cite{bertone2}: a cone of half-aperture 4'' pointed at the GC and $\nu=0.408$ GHz (case 1), a region with angles from the GC between 5' and 10' and $\nu=0.327$ GHz (case 2), and finally a cone of half-aperture 13.5' pointed at the GC and $\nu=0.327$ GHz (case 3). We use the measured fluxes quoted in \cite{bertone2}. It turns out that, independently of the annihilation channel, case 1 gives the most stringent bounds when using the NFW smooth profile suggested by Via Lactea II simulation. For the Aquarius simulation and its Einasto smooth profile, case 2 is the most constraining one.

\par In principle, for DM particles with Sommerfeld-enhanced cross sections, a full calculation of the synchrotron emission should include the enhancement $S(v)$ inside the integral in expression \eqref{nudwdnu}. Nevertheless, for our present proposes it suffices to put $\sigma_{ann} v \sim S(v_{\odot}) (\sigma_{ann} v)_0$ since the signal comes mainly from regions where $v\sim v_{\odot}$ $-$ recall that $v\sim v_{\odot}$ at $r=R_{acc}=0.04$ pc. In case 1, for instance, one is looking into a region of size $\sim 0.16$ pc around the GC. The region defined in case 2 encompasses distances of $\sim 10-20$ pc from the GC where $v<v_{\odot}$ and $S(v)\geq S(v_{\odot})$; hence, in this case, considering $\sigma_{ann} v \sim S(v_{\odot}) (\sigma_{ann} v)_0$ yields actually a lower bound on the radio flux.

\par
\begin{figure}
 \centering
 \includegraphics[width=7.5cm,height=7.5cm]{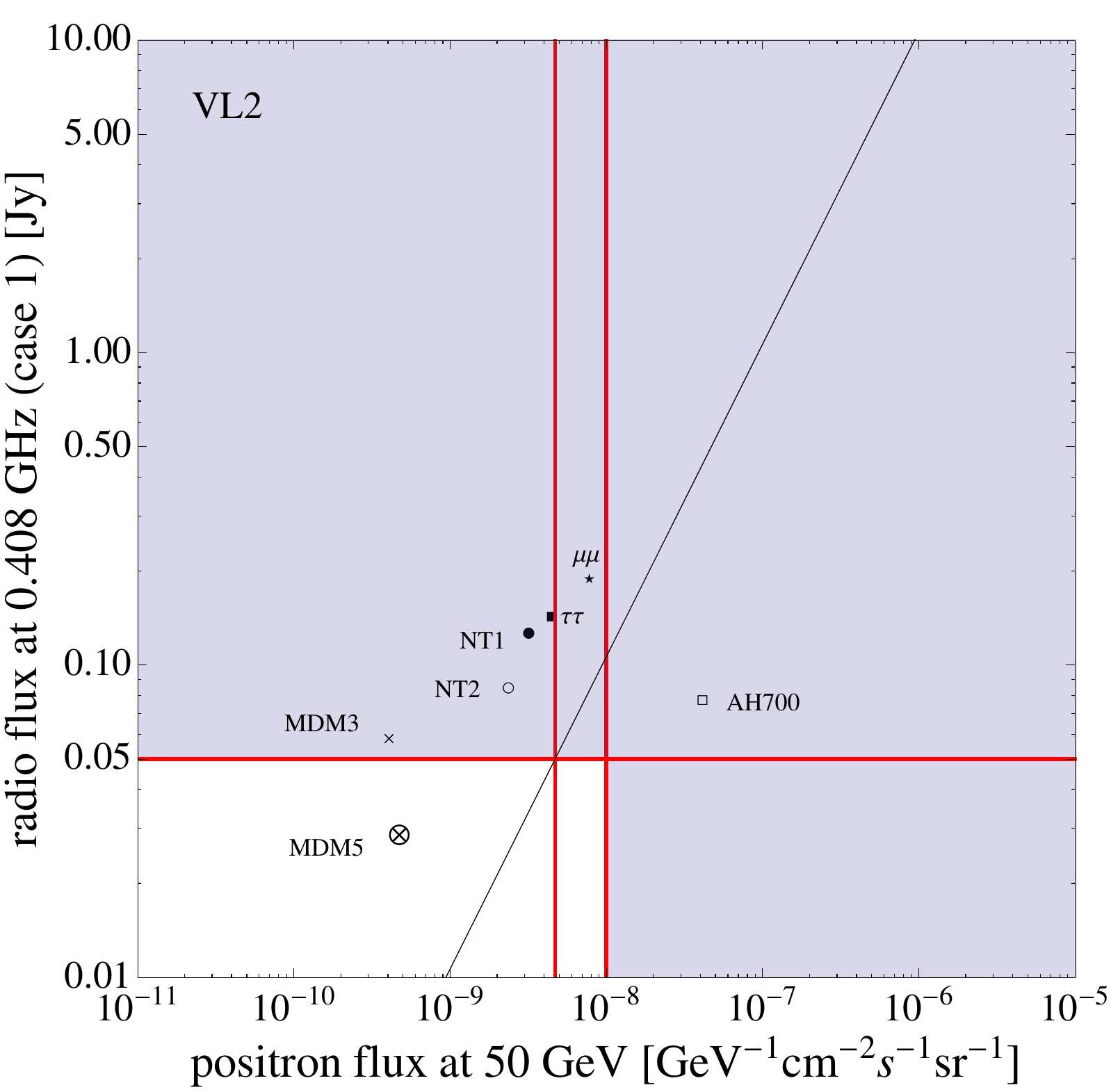}\\
 \includegraphics[width=7.5cm,height=7.5cm]{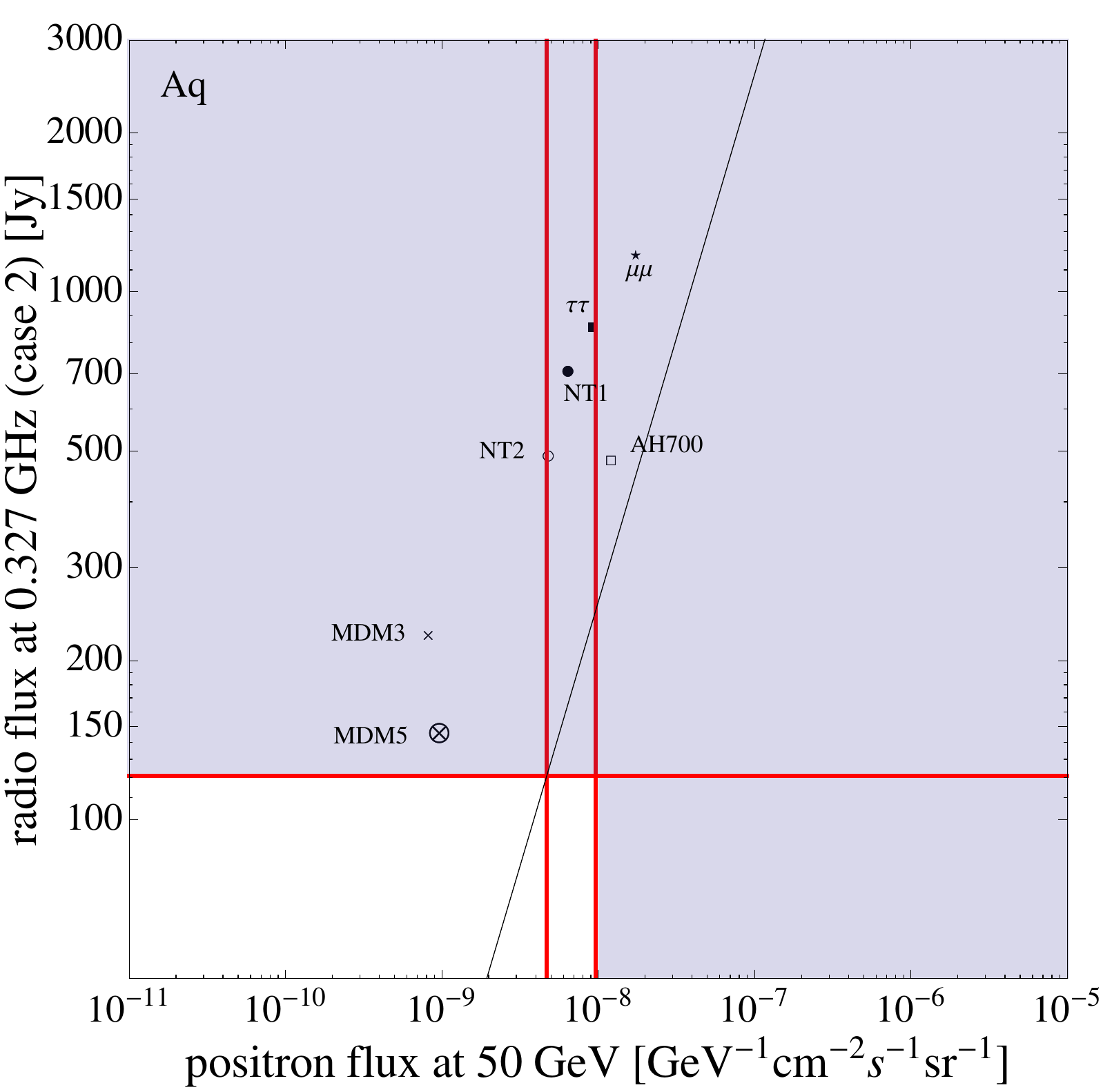}
 \caption{\fontsize{9}{9}\selectfont (Color online) The radio flux for case 1 (2) and Via Lactea II (Aquarius) parameters against the positron flux at 50 GeV already presented in figure \ref{fige+pbar}. The MED propagation is used and $M_{min}=10^{-6} \textrm{ M}_{\odot}$. The horizontal line indicates the measured flux (see \cite{bertone2} for details). Similarly to figure \ref{fige+pbar}, candidates lying above the diagonal line cannot be rescaled to explain PAMELA positron excess without violating radio bounds. Shaded regions are excluded by PAMELA (assuming the electron fluxes discussed in the text) and radio observations.}\label{figradio}
\end{figure}

 The results for the radio flux (case 1 for Via Lactea II and case 2 for Aquarius) are presented in figure \ref{figradio}, plotted against the corresponding positron fluxes at 50 GeV. In the case of Via Lactea II, we can see that most candidates seem to be at odds with radio observations even when a rescaling of $(\sigma_{ann}v)_0$ is applied to meet the positron excess. The situation for Aquarius (using case 2) is similar.

\section{Conclusions}
\par Table \ref{tabconc} summarizes our main results. There, we present for each model under consideration the maximum $(\sigma_{ann}v)_{0}$ allowed by the antiproton bound, the HESS measurement from the GC and radio observations. For antiprotons we conservatively use the largest value in equation \eqref{pamelap}. The most constraining of the three limits, i.e. the one yielding a minimal $(\sigma_{ann}v)_{0,max}$, is displayed in bold. We have disregarded here diffuse $\gamma$-rays as well as radio fluxes in the cases 2 and 3 (1 and 3) for Via Lactea II (Aquarius) since they give subdominant contraints. Furthermore, the $e^+$ column shows the value of $(\sigma_{ann}v)_{0}$ needed to meet the lowest positron flux in equation \eqref{pamelae}; these numbers are underlined only if allowed by the most stringent bound in bold. Notice that we apply this procedure to MDM3 and MDM5 even though minimal dark matter is a rather predictive scheme.

\begin{table*}[ht]
\centering
\fontsize{9}{9}\selectfont
\begin{tabular}{l|c|c|c|c||c|c|c|c}
\hline
\hline
 & \multicolumn{8}{c}{$(\sigma_{ann}v)_{0,max}/( 10^{-26} \textrm{cm}^{3}\textrm{s}^{-1})$}\\
\cline{2-9}
 & \multicolumn{4}{c||}{Via Lactea II} &\multicolumn{4}{c}{Aquarius} \\
\cline{2-9}
label & $e^{+}$ & $\bar{p}$ & $\gamma$ GC & radio (1) & $e^{+}$ & $\bar{p}$ & $\gamma$ GC & radio (2)\\ 
\hline
AH700 & $\underline{0.34}$ & $-$ & $49$  & $\bf 2.0$  & $1.2$ & $-$ & $17$  & $\bf0.76$   \\
NT1 & $4.4$ & $840$  &  $427$ & $\bf1.2$ & $2.2$ & $200$ & $136$ & $\bf0.51$   \\
NT2 & $6.0$ & $-$ & $238$ & $\bf1.8$ & $2.9 $ & $-$ & $76$ & $\bf0.74  $   \\
$\mu\mu$ & 1.8 & $-$ & 6.5 & $\bf0.80$ & 0.81 & $-$ & 2.1 & $\bf0.31$ \\
$\tau\tau$ & 3.1 & $-$ & 1.3 & $\bf 1.1$ & 1.5 & $-$ & $\bf 0.41$ & 0.43 \\
\hline
MDM3 & $12  $ & $7.9  $ & $ 9.4  $  & $\bf0.86$  & $5.7  $ & $ 1.9 $ & $3.0$ & $\bf0.54  $    \\
MDM5 & $9.9  $ & $69$ & $7.7  $ & $\bf1.7$  & $4.9  $ & $18$ & $2.5$ &  $\bf0.82  $   \\
\hline
\end{tabular}
\caption{\fontsize{9}{9}\selectfont The maximum allowed $(\sigma_{ann}v)_{0}$ by the antiproton bound, the $\gamma$-ray measurement from the GC and radio observations. The MED propagation is used and $M_{min}=10^{-6} \textrm{ M}_{\odot}$. The bold values represent the most constraining channel and the $e^+$ columns display the $(\sigma_{ann}v)_{0}$ needed to meet the positron excess. Underlined values manage to explain the positron excess while being allowed by our multi-messenger scheme of constraints.}\label{tabconc}
\end{table*}

\begin{figure*}[htp]
 \centering
 \includegraphics[width=7.5cm,height=7.5cm]{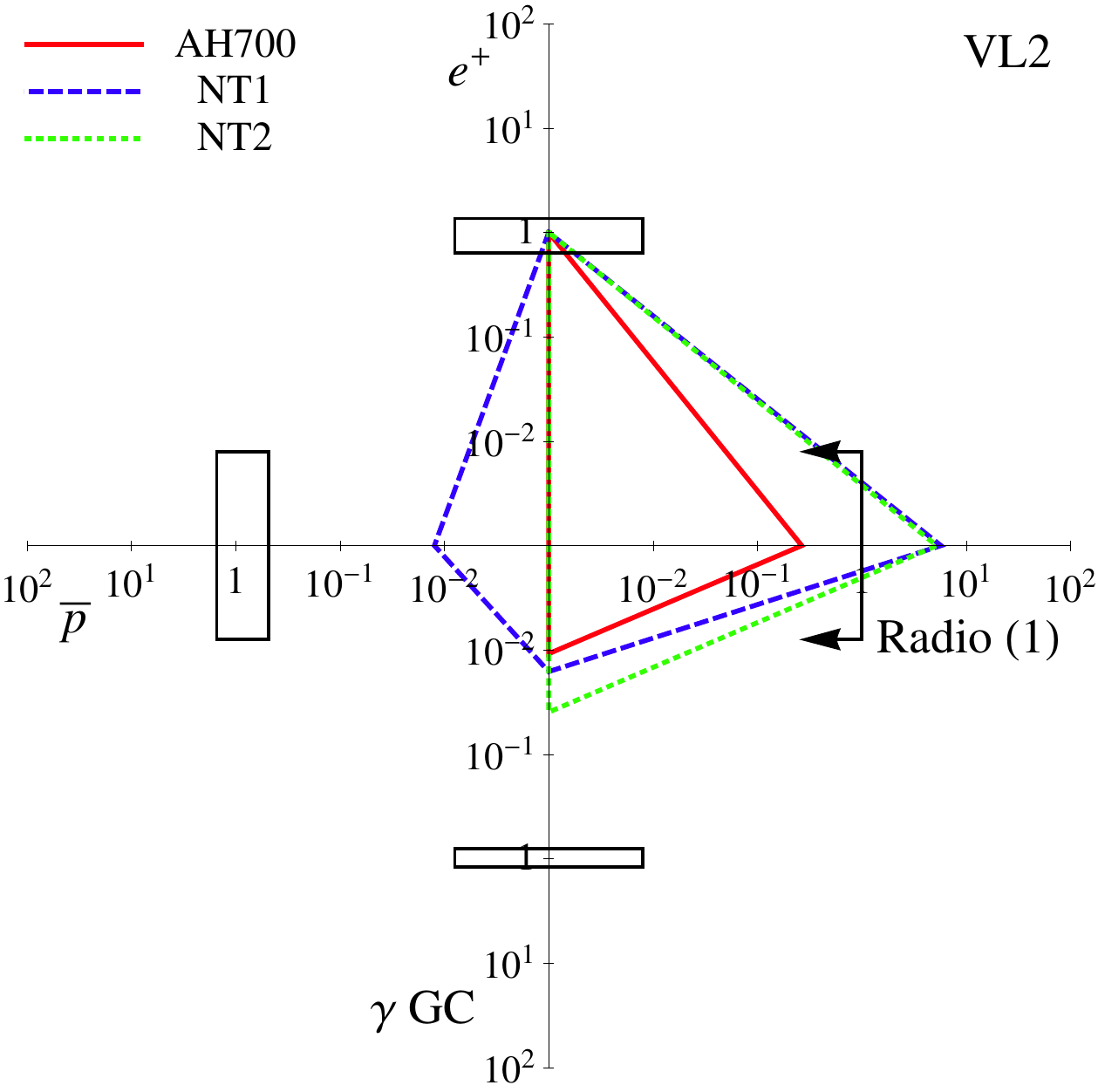}\hspace{0.1cm}
 \includegraphics[width=7.5cm,height=7.5cm]{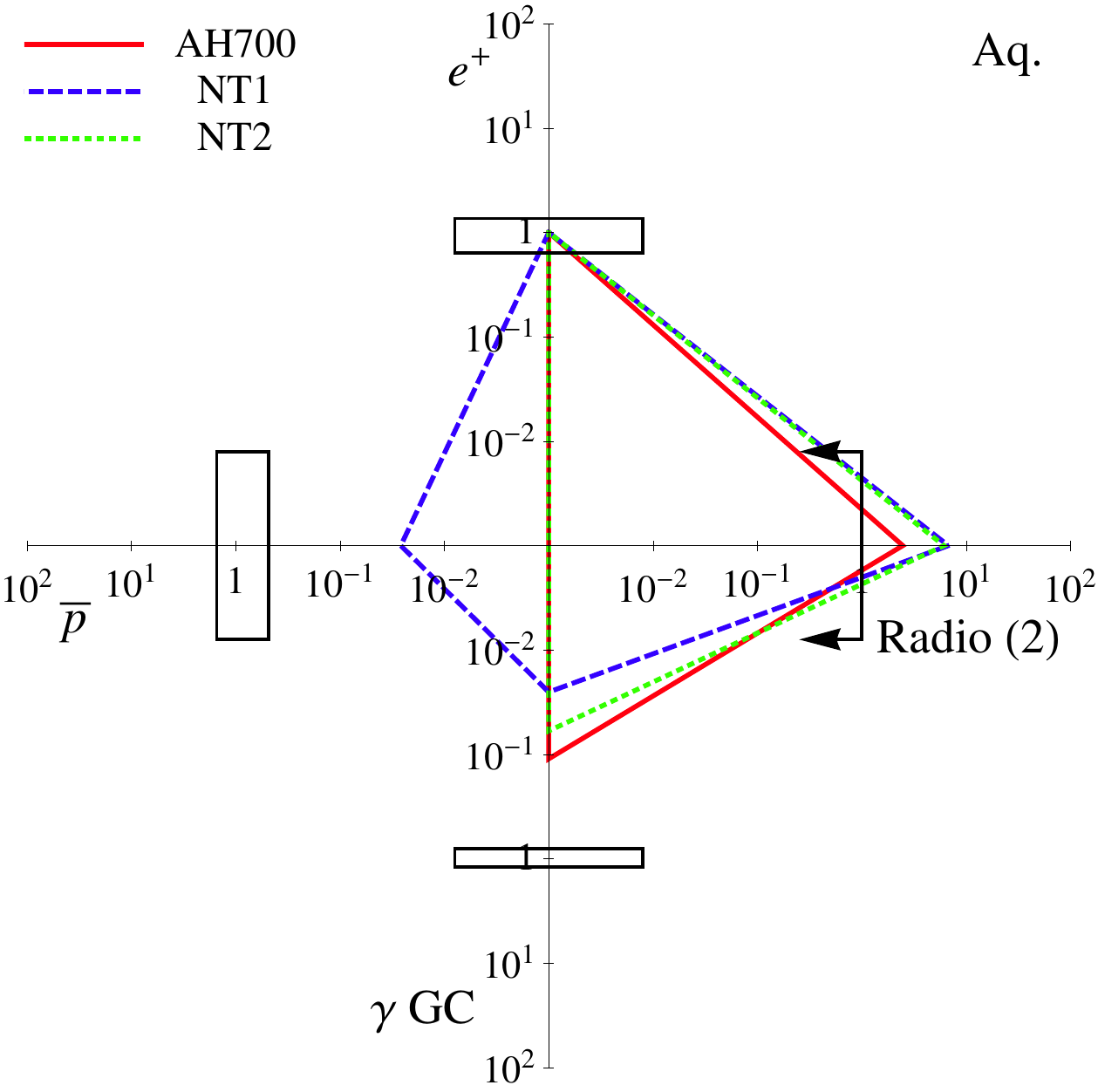}\\ \vspace{0.3cm}
 \includegraphics[width=7.5cm,height=7.5cm]{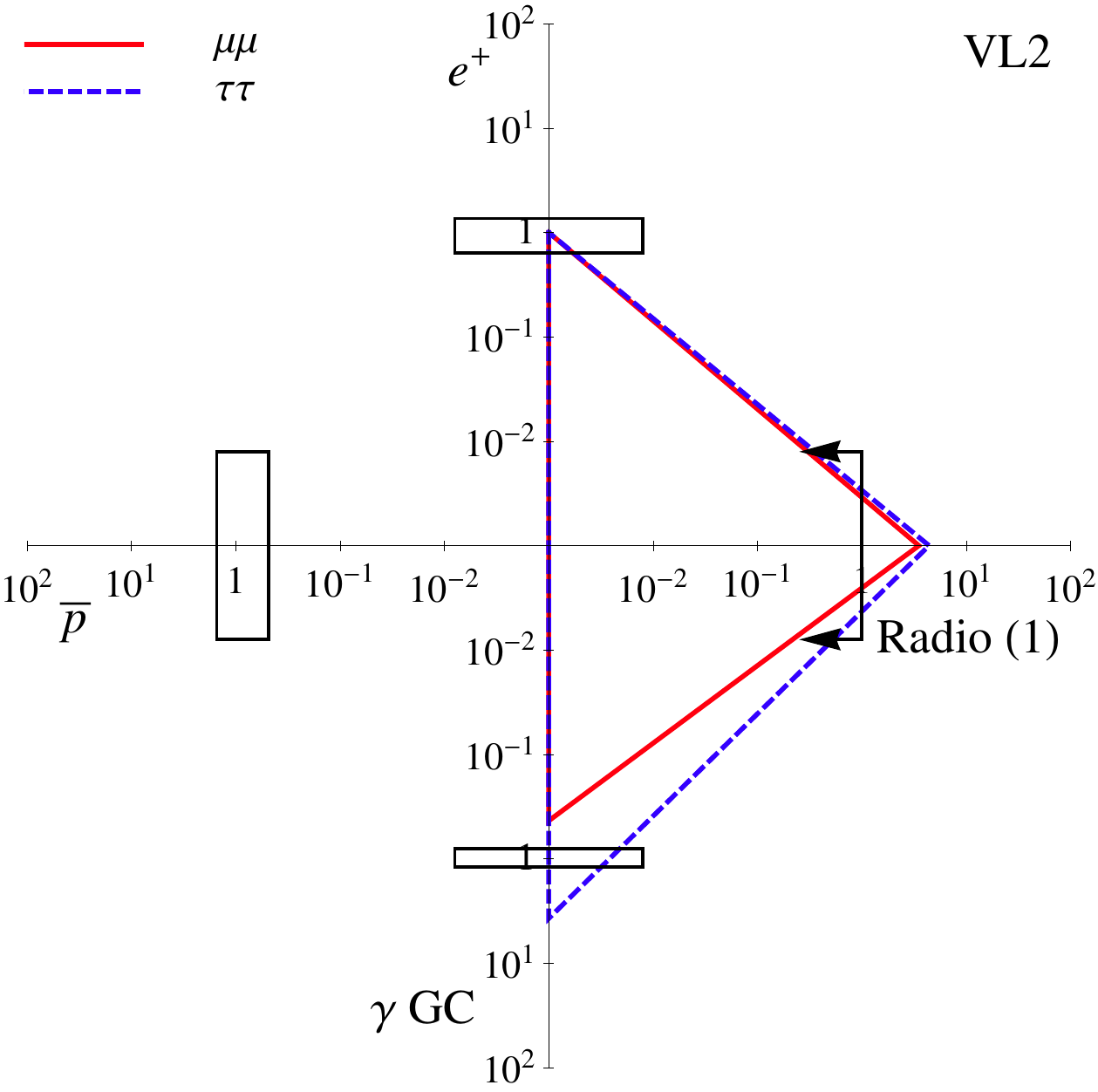}\hspace{0.1cm}
 \includegraphics[width=7.5cm,height=7.5cm]{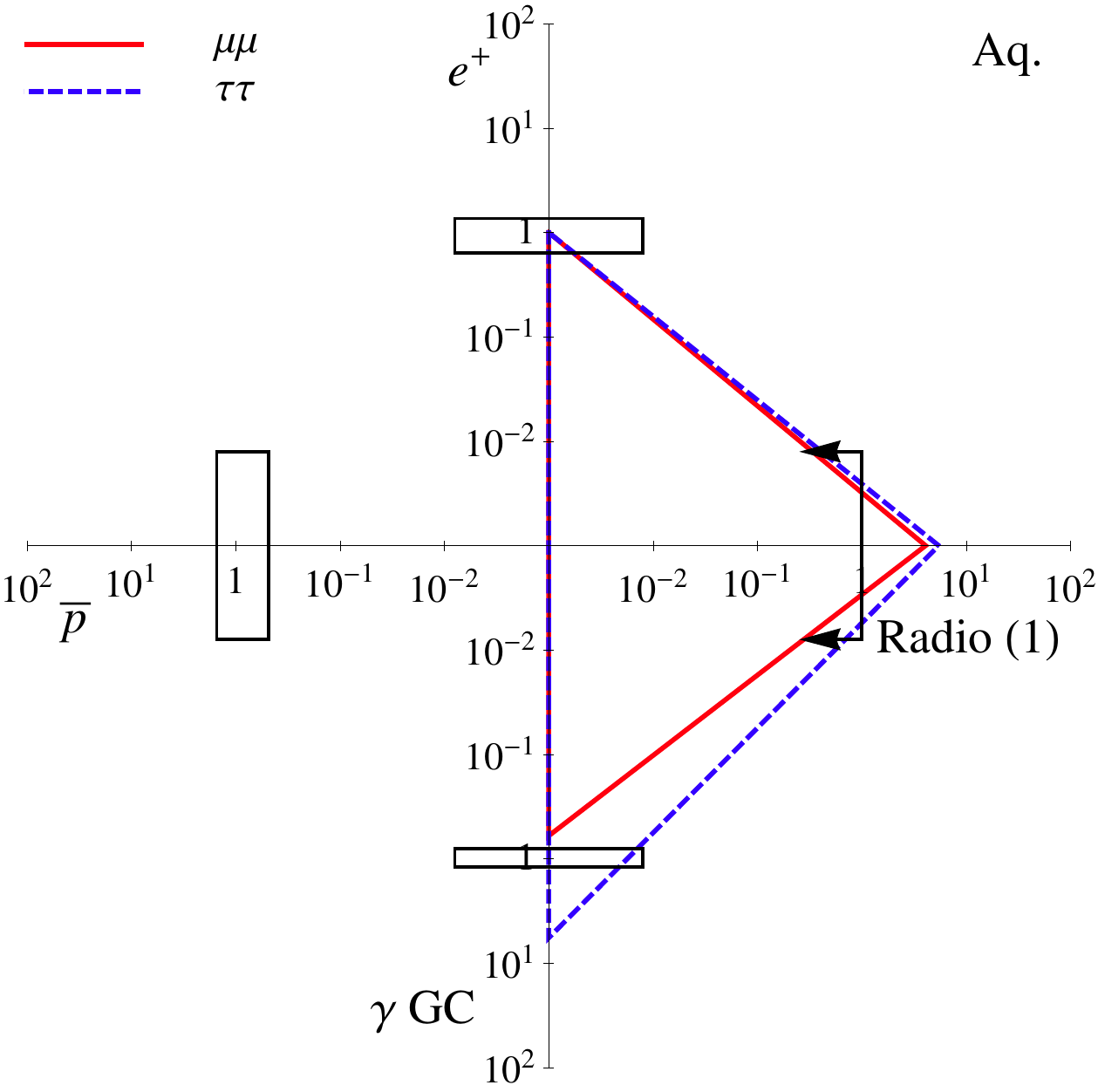}\\ \vspace{0.3cm}
 \includegraphics[width=7.5cm,height=7.5cm]{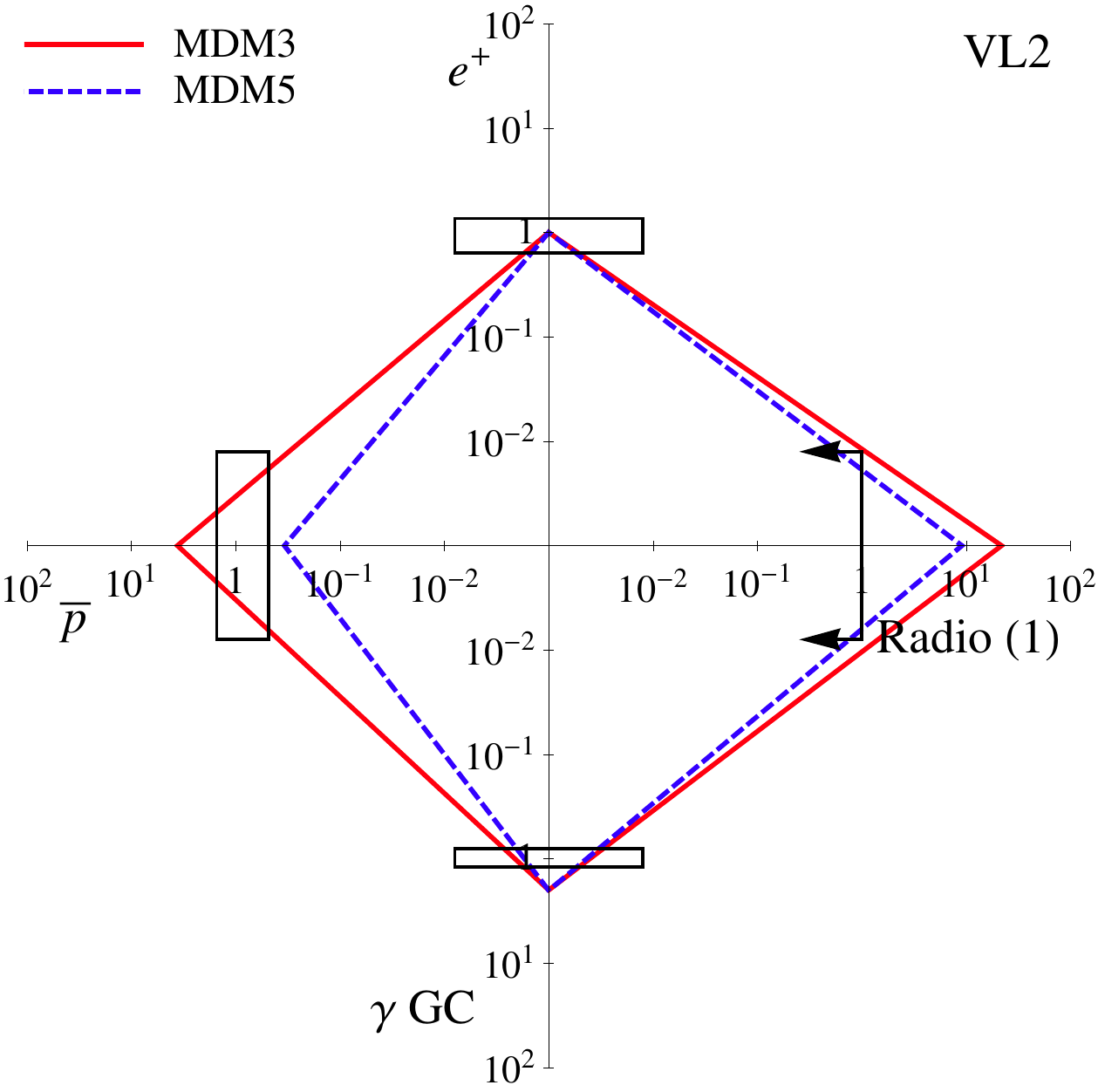}\hspace{0.1cm}
 \includegraphics[width=7.5cm,height=7.5cm]{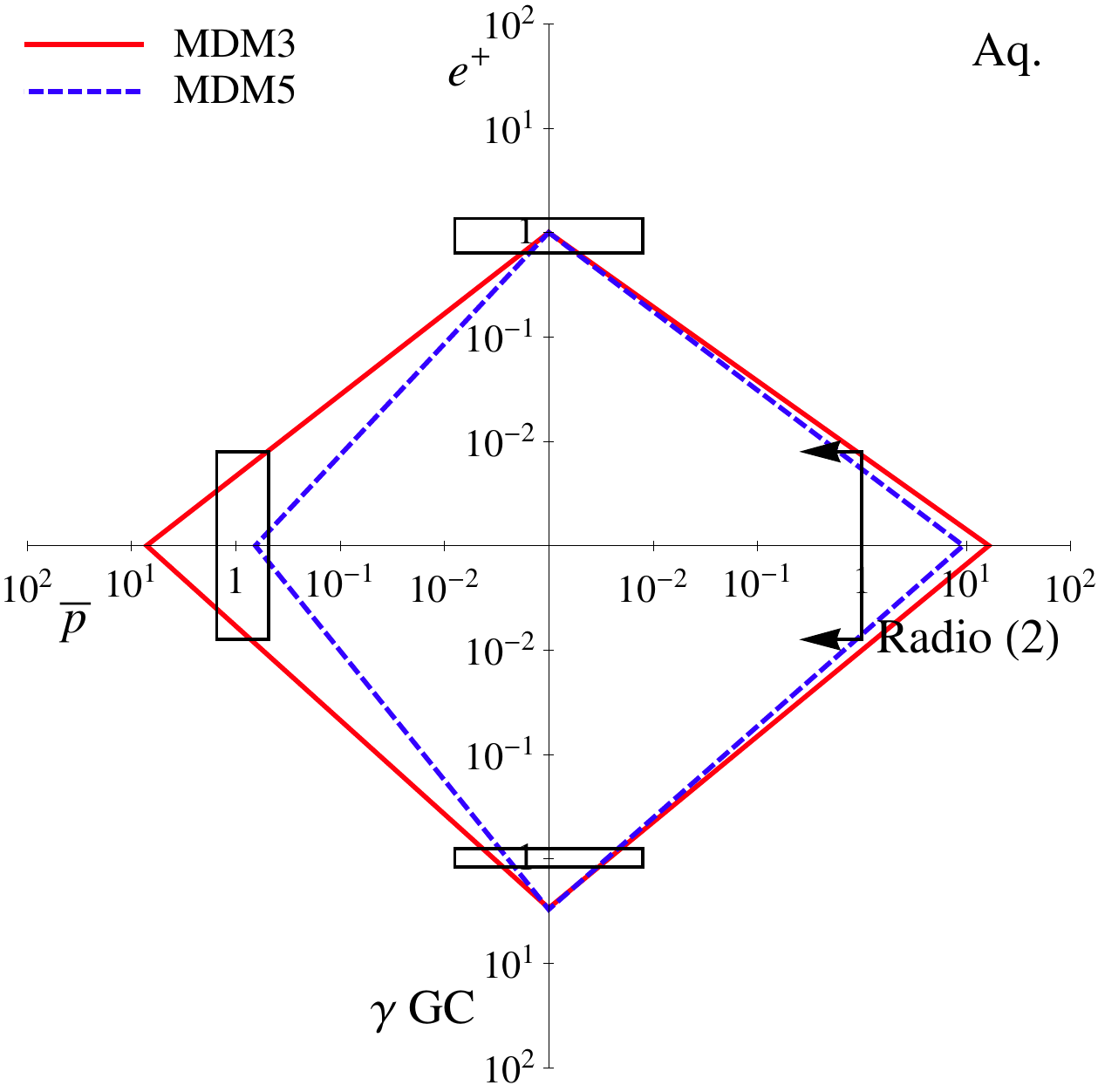}
\caption{\fontsize{9}{9}\selectfont (Color online) Multi-messenger bounds for several DM candidates. The MED propagation is used and $M_{min}=10^{-6} \textrm{ M}_{\odot}$. The $e^+$, $\bar{p}$, $\gamma$-ray, radio (cases 1 and 2) axes are normalised to $\phi_{e^+}(50\textrm{ GeV})=7.35\cdot 10^{-9} \textrm{ GeV}^{-1}\textrm{cm}^{-2}\textrm{s}^{-1}\textrm{sr}^{-1}$, $\phi_{\bar{p}}(50\textrm{ GeV})=5.2\cdot 10^{-9} \textrm{ GeV}^{-1}\textrm{cm}^{-2}\textrm{s}^{-1}\textrm{sr}^{-1}$, $\phi_{MW}(\psi=0^{\circ})=1.89\cdot10^{-11}\textrm{cm}^{-2}\textrm{s}^{-1}$, 0.05 Jy and 121 Jy, respectively. The boxes encompass the values in equations \eqref{pamelae} and \eqref{pamelap} for positrons and antiprotons, and a 20\% uncertainty on top of the HESS measurement for $\gamma$-rays. Notice that changing the value of $(\sigma_{ann}v)_{0}$ leads to an overall scaling of the polygon.}\label{fig4D}
\end{figure*}

\par Firstly, we immediately see from table \ref{tabconc} that radio observations are rather constraining w.r.t.~antiprotons or $\gamma$-rays for models trying to explain the positron excess. In the case of Via Lactea II DM distribution, just one candidate survives the studied bounds: AH700. There is still some tension with the data though. In fact, in order to fit the PAMELA data, this model needs to be rescaled down to $(\sigma_{ann}v)_{0}$ of $\sim 10^{-27}\textrm{ cm}^3\textrm{s}^{-1}$, which means either that such particles would overclose the universe in the standard thermal relic scenario, or that they are prevented to be the dominant dark matter component in the universe. 
As far as Aquarius is concerned, we identify no model that can evade all the implemented bounds.
%

\par In figure \ref{fig4D} we introduce a new method to visualise the multi-messenger constraints. We place each of the channels in a semi-axis and normalise it to the experimental limits. Since we are interested in the viability of the DM explanation of the positron excess, we only show models that are able to reproduce the observed PAMELA flux, and thus cross the ``up'' axis at 1. Configurations exceeding the boxes on other axes violate observational bounds and are therefore ruled out. Configurations not crossing the boxes are in principle viable, but one has to check then whether the cross section allows to achieve the appropriate relic abundance, whether the ``boost-factors'' are reasonable, and whether the model provides a good fit to {\it all}  PAMELA data. We stress that once the model is specified, the shape (angles and number of vertices) is fixed, and changing the cross-section corresponds to increasing or decreasing the overall size of the polygon. Note for instance that leptophilic configurations (first and second rows of figure \ref{fig4D}) have a different shape with respect to hadrophilic ones (third row).

\vspace{0.5cm}
\par To sum up, we have analysed the possibility that the models in table \ref{tab2}, recently suggested in the literature, may explain the PAMELA positron excess without violating bounds in the antiproton, $\gamma$-ray and radio channels. It turns out that $-$ even considering both substructure and Sommerfeld enhancement $-$ the candidates that provide a good fit to positron data, inevitably overproduce antiprotons, gamma-rays or radio emission. Our conclusions hold for the DM distributions from Via Lactea II and Aquarius simulations, the MED propagation model and $M_{min}=10^{-6}\text{ M}_{\odot}$.
As discussed in section \ref{secpbar}, modifying the propagation parameters does not change our main results. This is because the antiproton bound $-$ the most sensitive to propagation $-$ is subdominant. Thus, a different propagation model is not sufficient to reconcile the studied candidates with the observational constraints. A non-standard DM profile, and a non-standard magnetic field profile at the Galactic Centre, can in principle make theoretical models compatible with observations, at the expenses of introducing new ad-hoc hypotheses on these quantities. \\

\begin{acknowledgments}
We thank Marco Cirelli, Massimiliano Lattanzi and Marco Taoso for useful discussions. MP would like to acknowledge a grant from Funda\c{c}\~{a}o para a Ci\^encia e Tecnologia (Minist\'erio da Ci\^encia, Tecnologia e Ensino Superior) and thank the IDAPP network. This work was supported in part by the French ANR project ToolsDMColl,
BLAN07-2-194882.
\end{acknowledgments}


\end{document}